\documentclass[11pt]{article}
\usepackage{graphicx}
\usepackage[utf8]{inputenc} 

\usepackage[export]{adjustbox}
\pdfoutput=1

\usepackage{amsmath,amsfonts,amssymb,epsfig}
\usepackage{slashed}
\numberwithin{equation}{section}

\usepackage{graphicx}
\usepackage{cancel}
\usepackage{cite}
\usepackage{color}
\usepackage{listings}

\usepackage{xspace}

\lstdefinestyle{mathematica}{
        basicstyle=\ttfamily\mdseries,
	language=bash,
	frame=false,
	xleftmargin=.25in}   

\lstdefinestyle{terminal}{
	language=bash,
	frame=lines,
	xleftmargin=.5in,
        numbers=none}
        
\lstdefinestyle{file}{
        basicstyle=\ttfamily\mdseries,
	language=bash,
	frame=shadowbox,
        numbers=left,   
        numberstyle=\tiny}

\newcommand{\du}[3]{#1_{#2}^{\,\,\,#3}}
\newcommand{\nm}[2]{N^#1_{\,\,\,#2}}
\newcommand{\cnm}[2]{N_#1^{*\,\,#2}}

\newcommand{\SARAH}{{\tt SARAH}\xspace}
\newcommand{\SPheno}{{\tt SPheno}\xspace}
\newcommand{\smh}{{\tt SMH}\xspace}

\newcommand{\exclude}[1]{}
\def\nn{\nonumber}

\def\Mm{M_M}

\def\beq{\begin{equation}}
\def\eeq{\end{equation}}
\def\bal{\begin{align}}
\def\eal{\end{align}}

\def\lagr{{\cal L}}

\def\s2b{s_{2\beta}}
\def\c2b{c_{2\beta}}

\long\def\symbolfootnote[#1]#2{\begingroup%
\def\thefootnote{\fnsymbol{footnote}}\footnote[#1]{#2}\endgroup}

\def\thv[#1,#2,#3]{\left( \begin{array}{c} #1 \\ #2 \\ #3 \end{array} \right)}
\def\twomat[#1,#2][#3,#4]{\left( \begin{array}{cc} #1 & #2 \\ #3 & #4 \end{array} \right)}
\def\threemat[#1,#2,#3][#4,#5,#6][#7,#8,#9]{\left( \begin{array}{ccc} #1 & #2 & #3\\ #4 & #5 & #6 \\ #7 & #8 & #9 \end{array} \right)}
\def\twovec[#1,#2]{\left( \begin{array}{c} #1  \\ #2 \end{array} \right)}

\def\ov{\overline}

\def\bB{\mathbf{B}}

\newcommand{\vtree}{V^{(0)}}

\newcommand{\vone}{V^{(1)}}

\DeclareMathOperator{\llog}{\overline{\text{log}}}
\DeclareMathOperator{\blog}{\overline{\text{log}}}

\newcommand{\re}{\mathrm{Re}}
\newcommand{\vev}[1]{\langle #1 \rangle}
\newcommand{\real}{\mathfrak{Re}}

\def\MS{\ensuremath{\overline{\mathrm{MS}}}\xspace}
\def\DR{\ensuremath{{\overline{\mathrm{DR}}^{\prime}}}\xspace}

\begin{document}

\begin{titlepage}

\begin{flushright}
KA-TP-23-2017
\end{flushright}
\begin{center}

\vspace{1cm}

{\LARGE \bf Supersymmetric and non-supersymmetric models\\\vspace{3pt} without catastrophic Goldstone bosons}

\vspace{1cm}

{\Large Johannes~Braathen,$^{\!\!\!\,a,b}$\symbolfootnote[1]{{\tt braathen@lpthe.jussieu.fr}}
Mark~D.~Goodsell$^{\,a,b}$\symbolfootnote[2]{{\tt goodsell@lpthe.jussieu.fr}}
and Florian~Staub$^{\,c,d}$\symbolfootnote[3]{{\tt florian.staub@kit.edu}}
}

\vspace*{5mm}

{\sl ${}^a$ LPTHE, UPMC Univ.~Paris 6, Sorbonne Universit\'es, 4 Place Jussieu, F-75252 Paris, France}
\vspace*{2mm}\\
{\sl ${}^b$ LPTHE, CNRS, 4 Place Jussieu, F-75252 Paris, France }
\vspace*{2mm}\\
{\sl ${}^c$ Institute for Theoretical Physics (ITP), Karlsruhe Institute of Technology,}\\[-0.2em]
{\it Engesserstra{\ss}e 7, D-76128 Karlsruhe, Germany}\\[0.2em]
{\sl ${}^d$ Institute for Nuclear Physics (IKP), Karlsruhe Institute of Technology,}\\[-0.2em]
{\it Hermann-von-Helmholtz-Platz 1, D-76344 Eggenstein-Leopoldshafen, Germany}
\end{center}

\vspace{0.7cm}

\abstract{The calculation of the Higgs mass in general renormalisable field theories has been plagued by the so-called ``Goldstone Boson Catastrophe,'' where light (would-be) Goldstone bosons give infra-red divergent loop integrals. In supersymmetric models, previous approaches included a workaround that ameliorated the problem for most, but not all, parameter space regions; while giving divergent results everywhere for non-supersymmetric models! We present an implementation of a general solution to the problem in the public code \SARAH, along with new calculations of some necessary loop integrals and generic expressions. We discuss the validation of our code in the Standard Model, where we find remarkable agreement with the known results. We then show new applications in Split SUSY, the NMSSM, the Two-Higgs-Doublet Model, and the Georgi-Machacek model. In particular, we take some first steps to exploring where the habit of using tree-level mass relations in non-supersymmetric models breaks down, and show that the loop corrections usually become very large well before naive perturbativity bounds are reached. 
}

\vfill

\end{titlepage}

\setcounter{footnote}{0}

\section{Introduction}
\label{SEC:intro}

The Large Hadron Collider has opened a new era of precision physics. Following the discovery of the Higgs, the measurement of its properties -- in particular its mass -- have now been performed with an astonishing precision. This is interesting because a precise determination of the Higgs mass is of crucial importance in understanding the fate of the Standard Model (it is used to calculate the Higgs quartic coupling, required to determine whether the electroweak vacuum is metastable) and is especially sensitive to new physics beyond the SM (BSM). This is particularly important in supersymmetric models, where there is a \emph{prediction} for the Higgs quartic coupling at tree level in terms of other fundamental parameters of the theory (notably the gauge couplings). There is therefore a long tradition of calculating higher order corrections to the Higgs mass which was founded at the beginning of the 90's when the dominant one-loop corrections in the minimal supersymmetric standard model (MSSM) were calculated \cite{Haber:1990aw,Okada:1990vk,Ellis:1991zd}. Nowadays, the dominant two- and even three-loop corrections are available for the MSSM in the gaugeless limit, with vanishing external momenta \cite{Hempfling:1993qq, Heinemeyer:1998jw, Heinemeyer:1998kz,
  Zhang:1998bm, Heinemeyer:1998np, Espinosa:1999zm, Espinosa:2000df,Degrassi:2001yf, Brignole:2001jy, Brignole:2002bz, Martin:2002iu,Martin:2002wn, Dedes:2003km, Heinemeyer:2004xw,Martin:2007pg, Harlander:2008ju,Kant:2010tf,Hollik:2014wea,Hollik:2014bua,Passehr:2017ufr} and the dominant momentum-dependent two-loop corrections were given in \cite{Martin:2004kr,Borowka:2014wla, Degrassi:2014pfa}.

However, all of these higher-order corrections bypass an intrinsic technical problem of divergences associated with would-be Goldstone bosons of the broken electroweak symmetry. Calculations beyond the Standard Model at two loops and higher have only been performed in Landau gauge in order to decouple ghosts and thus simplify the calculations; however, in this gauge the would-be Goldstone bosons are massless and lead to infra-red divergences. In the MSSM, the gaugeless limit avoids this by turning off the Goldstone boson\footnote{We shall drop the prefix ``would-be'' from now on; although in the gaugeless limit this distinction is irrelevant because they become physical Goldstone bosons.} couplings to the Higgs; and the other (momentum-dependent) calculations that have been performed beyond this limit only consider the sector of the theory without the Goldstones.  

However, as soon as one considers non-minimal supersymmetric models in which trilinear interactions of the Higgs superfields occur in the superpotential, the gaugeless limit no longer offers much protection against the problem, since the quartic coupling is not determined by the gauge couplings; and this is a generic feature of non-supersymmetric models (such as the Standard Model!). The so-called ``Goldstone Boson Catastrophe'' was noticed in the first attempt to go beyond the gaugeless limit in the MSSM at more than one loop \cite{Martin:2002wn}, and leads to divergent values for the Higgs mass at two loops and beyond -- it can in fact be a complete obstacle to a precise calculation. 

Recently a solution was proposed in the context of the Standard Model \cite{Martin:2014bca,Elias-Miro:2014pca} (see also \cite{Andreassen:2014eha,Espinosa:2016nld,Espinosa:2016uaw} for recent related work) and then extended to the MSSM \cite{Kumar:2016ltb} which involved resumming (a subset of) the Goldstone boson propagators. An alternative for the Standard Model based on the 2PI effective action was proposed in \cite{Pilaftsis:2013xna,Pilaftsis:2015cka,Pilaftsis:2015bbs}, where essentially all particle propagators are resummed. However, both of these approaches are difficult to generalise. Instead, in \cite{Braathen:2016cqe} a general procedure was developed to cure this problem in two-loop Higgs mass calculations, based on setting the Goldstone boson propagators on-shell, which provided a complete set of modified loop functions for the tadpoles and self-energies that were finite. Thus, combining the results of \cite{Braathen:2016cqe} with those of \cite{Martin:2003it,Martin:2003qz,Martin:2005eg,Goodsell:2015ira} which provide fully generic expressions for the two-loop corrections to real scalar masses in supersymmetric and non-supersymmetric models, all ingredients are present to calculate Higgs masses in {\it any} renormalisable model. 

The generic expressions of \cite{Martin:2003it,Martin:2003qz,Martin:2005eg,Goodsell:2015ira} are already used by the Mathematica package \SARAH \cite{Staub:2008uz,Staub:2009bi,Staub:2010jh,Staub:2012pb,Staub:2013tta,Staub:2015kfa} to calculate in combination with \SPheno \cite{Porod:2003um,Porod:2011nf} the Higgs masses in supersymmetric models at the two-loop level \cite{Goodsell:2014bna,Goodsell:2015ira,Goodsell:2016udb}. Up to now, the workaround for the Goldstone boson catastrophe in this setup was to introduce finite masses for the electroweak Goldstones by dropping the $D$-terms in the mass matrices. However, there were many regions of parameter space where the divergences reappeared (see e.g. \cite{Goodsell:2015ura,Benakli:2016ybe,Athron:2016fuq}) and this does not work at all for non-supersymmetric models, which have no $D$-term potential! Therefore, to perform this work we have implemented the results of \cite{Braathen:2016cqe}, in addition to filling some additional technical gaps which we describe here in section \ref{SEC:GBC} and the appendices; in particular, we complete the basis of required loop functions. The new version of \SARAH {\tt\ 4.12.0} therefore now offers the possibility to calculate two-loop masses for neutral scalars in non-supersymmetric models, as well as substantially improving the calculation in supersymmetric ones. As the only non-supersymmetric model for which comparable results exist is the Standard Model, in section \ref{SEC:SM} we compare our new calculation against the public code \smh \cite{Martin:2014cxa} and the results of \cite{Buttazzo:2013uya}, finding excellent agreement (even if our results do not include all of the contributions included in those references). We then illustrate our new routines by computing some new results in Split SUSY in section \ref{SEC:SPLIT}. On the other hand, in section \ref{sec:SUSY} we show how our new approach improves our previous calculation for supersymmetric models through the example of the NMSSM, for which our results should now be considered state of the art.

Momentum-independent renormalisation schemes are the most convenient choices for applying to a large variety of models, and so all mass calculations in \SARAH are performed in the \MS or \DR scheme. In contrast, on-shell schemes might offer some model dependent advantages. This is for instance the case in supersymmetric models with Dirac gauginos and a large mass splitting between the stops and the gluino. It has been shown that in this case an on-shell scheme leads to an improved convergence of the perturbative series \cite{Braathen:2016mmb}. It is also very useful often if a \DR and on-shell calculation exists for the same supersymmetric model: the difference between the results can be used as estimate of the missing higher-order corrections; this can now be done for the MSSM and certain classes of NMSSM and Dirac gaugino contributions. On the other hand, there has been hardly any discussion in the literature about radiative corrections to Higgs masses in non-supersymmetric BSM models. One reason for this, besides the technical hurdles, is that the additional freedom in non-supersymmetric models introduces a large number of free parameters, i.e. in some cases it might be possible to absorb any finite correction in the scalar sector into the counter-terms of these parameters. Thus, it is often implicitly assumed that the masses, but also the mixing angles, in the extended Higgs sector in BSM could be kept at their tree-level values. However, this is fraught with danger: (i) not all non-supersymmetric models really have a sufficiently large number of free parameters to absorb all radiative corrections. This is for instance the case in the Georgi-Machacek model. (ii) if a low-energy model is combined with an explicit UV completion (such as a GUT theory), the freedom to adjust the couplings is usually lost. (iii) using masses instead of couplings as input hides the presence of huge or even non-perturbative quartic couplings. (iv) even if parameters are checked with respect to simple limits such as $\lambda < 4\pi$ or tree-level unitarity bounds, this does not guarantee that the considered parameter point is perturbative or that strongly coupled effects do not appear at lower energies than can be explored at the LHC. Partly motivated by the growing interest in exploring quantum corrections to non-supersymmetric models, here in sections \ref{SEC:THDM} and \ref{SEC:GM} we explore the corrections to the Two-Higgs-Doublet Model (2HDM) and Georgi-Machacek model (GM), drawing attention to the fact that the corrections pass out of control well before the naive perturbativity or unitarity bounds. 

Finally, an \MS calculation has the advantage that it can give an impression of the size of the theoretical uncertainty by varying the renormalisation scale. Moreover, to obtain more reliable results for the vacuum stability by considering the renormalisation group equation (RGE) improved effective potential, a translation of masses into \MS parameters is necessary. We show in this work how these aspects can be analysed in non-supersymmetric models with the new calculation available now in \SARAH.

\section{The Goldstone Boson Catastrophe and its solutions}
\label{SEC:GBC}
To calculate the Higgs boson masses in general field theories we require the tadpole diagrams and self-energies. Expressions for the former were given in \cite{Goodsell:2015ira}, which were derived from the general expression for the effective potential in the Landau gauge, given in \cite{Martin:2001vx}. Hence we must also use the self-energies in the Landau gauge; these were given in \cite{Martin:2003it} up to order $g^2$ in the gauge couplings, and so we restrict ourselves to the ``gaugeless limit'' where we ignore the contributions of broken gauge groups. This has a number of advantages, chiefly simplicity and speed of the calculation; but also the fact that we can compute the one-loop corrections in any gauge desired. Once we have dropped the electroweak contributions, it is also tempting to disregard the momentum-dependence of the loop functions, which is typically estimated to contribute at the same order (and indeed is so for the MSSM \cite{Borowka:2014wla, Degrassi:2014pfa}) -- hence the popularity of calculations in the effective potential approach.

However, calculations in the Landau gauge/gaugeless limit suffer from the ``Goldstone Boson Catastrophe'', where the Goldstone bosons lead to ill-defined or divergent loop functions. Let us define the scalar potential in terms of real scalar fields $\varphi^0_i$ and their fluctuations around expectation values $v_i$ such that $\varphi^0_i \equiv v_i + \phi^0_i$ (not necessarily mass diagonal):
\begin{align}
\label{scalar_pot}
 \vtree(\{\varphi_i^0\})=& \frac{1}{2} m^2_{0,ij} \varphi_i^0 \varphi_j^0 + \frac{1}{6}\lambda_0^{ijk}\varphi_i^0\varphi_j^0\varphi_k^0+\frac{1}{24}\lambda_0^{ijkl}\varphi_i^0\varphi_j^0\varphi_k^0\varphi_l^0 \nn\\
=&\vtree(v_i) + t^i \phi_i^0 +\frac{1}{2} m_{ij}^2 \phi_i^0\phi_j^0+\frac{1}{6}\hat{\lambda}_0^{ijk}\phi_i^0\phi_j^0\phi_k^0+\frac{1}{24}\hat{\lambda}_0^{ijkl}\phi_i^0\phi_j^0\phi_k^0\phi_l^0,
\end{align}
where $t^i$ are tadpoles. Since we define the VEVs to be exact, we must have
\begin{align}
t^i + \frac{\partial \Delta V (\{m_{ij}^2\})}{\partial \phi^0_i}\bigg|_{\phi^0_i = 0} =&0.
\end{align}
By defining the potential in terms of fluctuations, we have the \MS/\DR masses squared $m_{ij}^2$ for all the scalars in the theory, and these are the values that enter the loop functions. However, the tadpoles are functions of the masses:
\begin{align}
t^i =& m^2_{0,ij} v_j +  \frac{1}{2}\lambda_0^{ijk}v_j v_k+\frac{1}{6}\lambda_0^{ijkl} v_j v_k v_l
\end{align} 
and so, since these need to be adjusted loop order by order, we must choose some parameters to vary -- and the standard choice is the mass-squared parameters, because in this way the couplings are unaffected. So then we define
\begin{align}
m_{ij}^2 \equiv \hat{m}_{0,ij}^2 + \Delta_{ij}
\end{align}
where $\hat{m}_{0,ij}^2 $ is the value without loop corrections (so $t^i = 0$)
and $m_{ij}^2$ satisfies the full tadpole equations:
\begin{align}
m_{ij}^2 v_j = \hat{m}_{0,ij}^2 v_j - \frac{\partial \Delta V (\{m_{ij}^2\})}{\partial \phi^0_i}\bigg|_{\phi^0_i = 0} ,
\label{EQ:masssqeq}
\end{align}
where $\Delta V$ consists of the loop corrections to the effective potential, and we have written explicitly its dependence on the parameters $\{m_{ij}^2\}$. 
The Goldstone Boson Catastrophe appears because the mass-squared parameter(s) of the Goldstone boson(s) in the Lagrangian is(are) zero at tree-level, but non-zero once we take into account the loop corrections to the potential. Then at two loops and higher we must calculate loop corrections with a small and/or negative mass-squared parameter, which leads to large logarithms and/or phases. 

In the context of the Standard Model \cite{Martin:2014bca,Elias-Miro:2014pca} (see also \cite{Andreassen:2014eha,Espinosa:2016nld,Espinosa:2016uaw}) and MSSM \cite{Kumar:2016ltb} it was suggested that resumming (a subset of) the Goldstone boson self-energies would cure divergences in the tadpole diagrams; and including external momenta in the self-energies would also be required to cure divergences there. Alternatively, refs.~ \cite{Pilaftsis:2013xna,Pilaftsis:2015cka,Pilaftsis:2015bbs} proposed using the (symmetry-improved) two-particle-irreducible potential to cure the problem in the Standard Model, which provides a consistent theoretical underpinning but unfortunately is particularly difficult to generalise. In the following we shall describe our previous approaches to the problem and the new results and implementation in \SARAH.

\subsection{Previous approaches in SARAH}

Up until now, in \SARAH the catastrophe appeared in an even more acute form because all of the one- and two-loop tadpoles and self-energies are computed using the tree-level masses in the loops, so without a solution to the problem, the Goldstone bosons are massless and cause several loop functions to diverge. However, for supersymmetric models the original workaround implemented in \cite{Goodsell:2014bna,Goodsell:2015ira} and explored in more detail in \cite{Goodsell:2016udb} relies on the fact that that the electroweak gauge couplings appear in the D-term potential.\footnote{Indeed, the gaugeless limit (turning off the electroweak gauge couplings) completely cures the problem in the MSSM by eliminating all of the Goldstone boson couplings to the Higgs.} We therefore used the tree-level parameters that are solutions of the full tree-level tadpole equations including the electroweak couplings to calculate the tree-level masses (but set the electroweak gauge couplings to zero in the mass matrices) used in the two-loop routines' loop functions. In other words, the masses in the loop functions are not at the minimum of the potential, and are typically tachyonic\footnote{Since the mass was tachyonic and generally not small, we then neglected the imaginary part of the self energies/tadpoles.}, with a size of order the electroweak scale. Since we are neglecting two-loop corrections proportional to these couplings, this error is acceptable. On the other hand, for models beyond the MSSM (in particular, the NMSSM) there are typically regions of the parameter space where the Higgs sector masses still pass near to zero and cause the loop functions to diverge; for example such problems were observed in \cite{Goodsell:2015ura,Benakli:2016ybe,Athron:2016fuq}.

A more recent approach was to introduce regulator masses. All scalar masses in the two-loop routines which are below a certain threshold are set in terms of the renormalisation scale $Q$ and a constant $R$:
\begin{equation}
m^2_{S,\rm min} = R Q^2 
\end{equation}
This approach was introduced in \SARAH to stabilise cases in which the $D$-term approach fails. This could either be, as demonstrated in an example in sec.~\ref{sec:SUSY}, if other scalars artificially become very light, or if the supersymmetric scale is much higher than the electroweak scale. However, in contrast to the D-term solution, this approach violates the symmetries of the theory and can lead to non-zero masses for Goldstone bosons. Furthermore, there is no a priori indication for the optimal size of $R$; too large and the Goldstone/Higgs contributions are suppressed (because logarithmic contributions including them are artificially reduced), too small and the results become numerically unstable, and the user must use trial and error. Finally, it implicitly assumes that the corrections coming from the Higgs/Goldstone bosons to the Higgs mass are small (so that modifying them is benign). This is not a good approximation in many non-SUSY models, and for this reason the newly implemented solution described in the next subsection allows non-SUSY models to be studied accurately for the first time.

\subsection{On-shell Goldstone bosons and consistent tadpole solutions}
\label{SEC:Consistent}

In \cite{Braathen:2016cqe} a genuine solution was presented for \emph{generic} field theories: we should treat the Goldstone boson mass as an on-shell parameter. A set of modified expressions for tadpoles and self-energies were given -- indeed, it was shown that there were a class of loop diagrams that were not made finite purely by including external momenta. In addition, expressions for the ``consistent solution'' of the tadpole equations were given. These two results are closely related, as we shall elaborate a little here.

If we take the Goldstone boson mass on-shell, as proposed in  \cite{Braathen:2016cqe}, then we have two possible ways of calculating the resulting tadpoles and self-energies, which differ in terms of how we solve the tadpole equations. The choice arises because the mass parameters $m_{ij}^2$ appear on both the left-  and right-hand sides of equation (\ref{EQ:masssqeq}), so we can:
\begin{enumerate}
\item Numerically solve equation (\ref{EQ:masssqeq}) to find the $m_{ij}^2$ \emph{exactly}.
\item Perturbatively expand the $m_{ij}^2$ so that 
$$m_{ij}^2 = \hat{m}_{0,ij}^2 + \delta^{(1)} m_{ij}^2 + \delta^{(2)} m_{ij}^2 + ...$$
and solve for a given loop order. 
\end{enumerate}

Since the effective potential $\Delta V$ will only be computed to a given loop order, the two approaches are formally equivalent. 
For the first approach, 
in practice, this means that we must iteratively solve the tadpole equations; at each iteration we put $m_i^2 = R_{ki} R_{li} m_{kl}^2$
for the tree-level mass parameters, computing a new $R$ each time and therefore modifying the couplings, and then set the Goldstone boson mass to zero in the loop functions and compute the tadpole equations from the expressions in  \cite{Braathen:2016cqe}.  
We find in this case that the couplings are no longer guaranteed to satisfy certain relationships imposed by the broken symmetries; only the full on-shell amplitudes will satisfy the appropriate Slavnov-Taylor identities. This is only a problem for the coupling $\lambda^{GG'G''}$ between three Goldstone bosons, which is zero at tree-level and on-shell; because the parameter in the Lagrangian will in general obtain a small non-zero value (in theories with CP-violation) and yet leads to divergent Goldstone boson self-energies we must impose that this is also on-shell (i.e. zero). Since this coupling does not appear at one-loop in the calculation of the Higgs boson mass, taking this coupling to vanish causes no shift at two loops. 

On the other hand, if we want to calculate the Goldstone boson self-energy at two loops then we \emph{do} find a set of shifts when we take this coupling ``on-shell'': we would need to include the vertex corrections and define a set of shifted loop functions for those contributions (which, of course, only affect the self-energies). We shall return to this in future work.

Instead, in our implementation of the results of \cite{Braathen:2016cqe} in \SARAH we take the second approach in the list above: we expand $m_{ij}^2$ as a series in the couplings, and solve explicitly up to two loop order in one step without recursion. This was already proposed in \cite{Kumar:2016ltb} for the MSSM, and in \cite{Braathen:2016cqe} explicit formulae for the corrections to the tadpoles and masses with a so-called \emph{consistent tadpole solution} were given for the general case. 
Then we can calculate all of our loop functions using the masses ($\hat{m}_{0,ij}^2$)  and couplings in the tree-level Lagrangian, and shifts $\Delta_{ij}$. 

However, here we shall also generalise a little the expressions given in \cite{Braathen:2016cqe}: we shall allow $\Delta_{ij}$ to be an implicit function of the tadpole shifts, rather than explicitly assuming $\Delta_{ij} = - \delta_{ij} \frac{1}{v_i} \frac{\partial \Delta V}{\partial \phi^0_i}\bigg|_{\phi^0_i = 0};$ indeed, this equation fails for pseudoscalars, for example. Instead we solve the tadpole equations for some variables $\{x_i\}$ with 
\begin{equation}
x_i = c_{0,i} + c_{ij} \times \frac{\partial \Delta V}{\partial \phi_j^0}
\label{EQ:tadpoleshift}
\end{equation}
then 
\begin{align}
\Delta_{ij} = & \sum_{k,l} \frac{\partial m^2_{ij}}{\partial x_k} c_{kl} \frac{\partial \Delta V}{\partial \phi_l^0}.
\end{align}
For example, in the Goldstone model of a single complex scalar $\Phi$ having potential
\begin{align}
V =& \mu^2 |\Phi|^2 + \lambda |\Phi|^4,
\end{align}
when $\Phi$ obtains a VEV it decomposes into a real scalar $h$ and a Goldstone boson $G$ as $\Phi = \frac{1}{\sqrt{2}} (v + h + iG).$ We solve the tadpole equations for the parameter $\mu^2$ so that
\begin{align}
\mu^2 + \lambda v^2 + \frac{1}{v} \frac{\Delta V}{\partial h} =0.
\end{align}
However, both the mass of the Goldstone boson and the Higgs are controlled by the $\mu^2$ parameter; 
\begin{align}
m_{hh}^2 = \mu^2 + 3 \lambda v^2, \qquad m_{GG}^2 = \mu^2 + \lambda v^2.
\end{align}
So in our notation, $x_h \rightarrow \mu^2 , c_{0,h} \rightarrow - \lambda v^2, c_{hh} \rightarrow - \frac{1}{v}$ and so 
\begin{align}
\Delta_{hh} =& - \frac{1}{v} \frac{\Delta V}{\partial h} = \Delta_{GG}.
\end{align}
Substituting the $\Delta_{ij}$ into the one-loop tadpole and self-energy expressions then gives a set of two-loop shifts; for scalars these were given in equations ($5.2$) and ($5.3$) of  \cite{Braathen:2016cqe}. 

We have implemented this under the assumption that the  variables $\{x_i\}$ are dimensionful and there is no \emph{explicit} dependence of the trilinear/quartic couplings on them (only implicitly through the mixing matrices $R$); and also we assume that the \emph{fermion} mass matrices do not depend on these parameters. These assumptions are fulfilled e.g. for $\{m_{H_u}^2, m_{H_d}^2\}$ in the MSSM, but not  $\{\mu,B_\mu\}$ chosen as parameters to solve the tadpole equations. On the other hand, we give expressions for the shifts to the tadpoles and self-energies when fermion masses depend on the $\{x_i \}$ in appendix \ref{APP:Consistent}, and plan to implement these in future.

Now, since the Goldstone boson is massless at tree-level, this then means that we automatically have the Goldstone boson ``on-shell.'' This means that the ``on shell'' and ``consistent solution'' approaches are more closely related than first appears: since the Goldstone boson mass must be zero on-shell and we can identify the Goldstone boson eigenstates using a matrix $R_{kG}$ derived just from the broken symmetries (see e.g. \cite{Braathen:2016cqe}) then the on-shell condition becomes 
\begin{align}
\mathrm{det} (p^2 - m_{ij}^2 - \Pi_{ij} (p^2)) = 0 \rightarrow R_{kG} R_{lG} m_{kl}^2 + \Pi_{GG} (0) = 0.
\end{align}
and since $\hat{m}_{0,GG}^2 =0$ we have
\begin{align}
\delta m_G^2 =& - R_{kG} R_{lG} m_{kl}^2 = - \Pi_{GG} (0) \nn\\
\Delta_{GG} =& \,\delta m_G^2 + \mathcal{O}(2-\mathrm{loop}) \rightarrow \Delta_{GG} = - \Pi_{GG} (0) + \mathcal{O}(2-\mathrm{loop}),
\end{align}
i.e. the approach of adjusting the loop functions (as we do when setting the Goldstone boson on-shell) or defining a set of shifts to the tadpoles and self-energies involving $\Delta_{ij}$ should give the same result when we just consider the shifts to the Goldstone boson masses, even though the expressions look very different.

\subsection{A complete basis of loop functions and the implementation in \SARAH}

For the evaluation of tadpoles and self-energies \cite{Braathen:2016cqe} proposed a ``generalised effective potential limit,'' where the self-energies are expanded in $s = - p^2 $ ($=m^2$ on shell) and all terms of order $\mathcal{O} (s)$ are neglected (but crucially retaining terms that diverge at $s=0$). We therefore require the following basis of loop functions, where $\{x,y,z,u,v\} \ne 0$ are masses squared:
\begin{align}
\mathrm{Momentum\ independent:}\ & J(x), P_{SS} (x,y), P_{SS} (0,y),I(x,y,z), I(0,y,z),  I(0,0,z), \nn\\
& U_0 (x,y,z,u),  U_0 (0,y,z,u), U_0 (x,y,0,u),U_0 (0,y,0,u), U_0 (x,y,0,0),U_0 (0,y,0,0),\nn\\
& M_0 (x,y,z,u,v),M_0 (0,y,z,u,v), M_0 (0,0,z,u,v),M_0 (0,0,0,u,v),\nn\\
& \tilde{V}(x,y,z). \nn\\
\mathrm{Momentum\ dependent:}\  & B(0,0),\nn\\
& M(x,0,0,0,0),M(0,y,0,u,v), M(0,0,0,u,v),M(0,0,0,0,v) \nn\\
& U(0,0,x,y), U(0,0,0,y)\nn\\
&\tilde{V}(0,y,z),\tilde{V}(0,0,z). 
\end{align}
All of these functions are implicitly dependent on the renormalisation scale $Q$, typically containing factors of $\blog x \equiv \log (x/Q^2)$. Expressions for all of these functions expanded up to $\mathcal{O}(1)$ in the external momenta (or the reference for them) were given in \cite{Braathen:2016cqe}. However, the functions $\tilde{V}(0,y,z), \tilde{V}(0,0,z)$ were given in terms of the regularised function $\ov{V}(u,v,y,z)$ defined in \cite{Martin:2003it}; unfortunately, however, no closed-form expression for this function was available, nor is it straightforward to evaluate it simply using the numerical package {\tt TSIL} \cite{Martin:2005qm}. Hence in appendix \ref{SEC:loopfn} we derive expressions for this function -- first with full momentum dependence, and then expanded up to $\mathcal{O}(1)$ in the external momenta.

In our practical implementation in \SARAH we have extended the available routines for calculating two-loop integrals with the missing ingredients to address the Goldstone boson catastrophe. Moreover, there are three loop functions involving fermions and gauge bosons which needed modification for the $\ov{\mathrm{MS}}$ scheme as used for non-supersymmetric models, as compared to the \DR for supersymmetric models; the tadpole and self-energies contain
\begin{align}
\frac{\partial \hat{V}^{(2)}}{\partial \phi_r^0} \supset& R_{rp} T_{FV}^p, \nn\\
T_{FV}^p  =& \,g^2  d(I) C(I) \mathrm{Re}( M_{I I'} y^{II'r}  ) \times \left(\frac{1}{2} F_{FV}^\prime(x)\right), \nn\\
\Pi_{ij}^{(2)} \supset \Pi_{ij}^{FV} =& \,g^2 d(K) C(K) \big[\mathrm{Re}(y^{iKL} y_{jKL}) G_{FF} (m_K^2,m_L^2)\nn\\
 &+ \mathrm{Re}(y^{ i KL} y^{j K'L'}M_{KK'} M_{LL'}  ) G_{\ov{FF}} (m_K^2,m_L^2)\big],
\end{align} 
where $\hat{V}^{(2)} $ is the two-loop contribution to the effective potential, $d(I), C(I)$ are the dimension and quadratic Casimirs of representation $I$ of the gauge group having coupling $g$, 
and the loop functions are:
\begin{align}
 \left(\frac{1}{2} F_{FV}^\prime(x)\right) =& 4x \bigg[ 6 - 7 \blog x  +3 \blog^2 x  + \delta_{\ov{\mathrm{MS}}} \big[ 2 \blog x -  1\big]\bigg], \nn\\
G_{FF} (x,y) =& G_{FF}^{\DR} (x,y) +2\delta_{\ov{MS}}  \bigg[ x + y + 2 J(x) + 2 J(y) - (x+y)\bigg( 2B(x,y) + x B(y,x') + y B(x,y')\bigg) \bigg] \nn\\ 
\underset{s\rightarrow 0}{\rightarrow}& \bigg[ 2 (x+y) [ 3U_0(x,y,x,0) + 3 U_0(x,y,y,0) + 5P_{SS}(x,y)] - 6I(x,x,0) - 6I(y,y,0) \nn\\
& + 10 J(x) + 10 J(y) - 16 (x+y) \bigg]+ 4\delta_{\ov{MS}}\bigg[x+y +J(x) + J(y) + (x+y) P_{SS}(x,y) \bigg],\nn\\
G_{\ov{FF}} (x,y) =& G_{\ov{FF}}^{\DR}(x,y) -4 \delta_{\ov{MS}} \bigg[ 2B(x,y)  +  y B(x,y') + x B(y,x')  \bigg]  \nn\\
\underset{s\rightarrow 0}{\rightarrow}& 4 \bigg( 3 U_0(x,y,x,0) + 3 U_0(x,y,y,0) +5 P_{SS}(x,y) -4 \bigg) +4 \delta_{\ov{MS}}  \bigg[  2P_{SS}(x,y) + 1 \bigg].
\end{align}
Here $ \delta_{\ov{\mathrm{MS}}} $ is one for \MS masses and zero for \DR.

In addition, routines to calculate the consistent tadpole solution are generated during the output of \SPheno code. This is fully automatised beginning with \SARAH version 4.12.0 and the user can obtain a \SPheno version for non-supersymmetric models as before -- with the difference that two-loop mass corrections are now included. We refer for more detailed explanations of how to use the code to the standard references such as \cite{Staub:2015kfa}. The only  
requirements  are recent versions of \SARAH and \SPheno which are available at {\tt www.hepforge.org}. 
\newpage

The new features can now be adjusted in the Block {\tt SPHENOINPUT} in the Les Houches input file: 
\begin{lstlisting}[style=file]
Block SPhenoInput #
...
  7  0   # Skip two loop masses: True/False
  8  3   # Choose two-loop method
150  1   # Use consistent tadpole solution: True/False
151  1   # Generalised effective potential calculations: True/False
410  0   # Regulator mass
\end{lstlisting}
Note that the solution to the Goldstone boson catastrophe exists only for the diagrammatic calculation (flag 8 $\to$ 3), but not for the effective potential calculations 
using numerical derivatives to obtain the tadpoles and self-energies (flag 8 $\to$ 1,2). By default, the new calculation is used now, but could be turned off if demanded (flag 151 $\to$ 0). In this case, it is usually necessary to include a non-zero regulator mass via flag 410 for non-supersymmetric models. In principle, there should not be any reason to revert to the old calculation with regulator masses except for double-checking the result. 

The consistent tadpole solution (described in the previous subsection) is turned off by default but can be turned on by setting flag 150 $\to$ 1. This is because, while strictly it is more accurate to include it, there is also the possibility of numerical instability if the shift in the tree-level mass parameters is large; for example, if the expectation values of some scalars are small (such as e.g. the neutral scalar of an electroweak triplet which must have a small expectation value from electroweak precision constraints) then the shift in the mass parameter can be much larger than the tree-level value and the perturbative solution fails. In such cases, it would be better to use a recursive approach which is currently not possible for the reasons given in section \ref{SEC:Consistent}.

\section{Standard Model}
\label{SEC:SM}
\subsection{A first comparison of our results with existing calculations}
Now that two-loop corrections to scalar masses are available in \SARAH, free of the Goldstone boson catastrophe, it is important to compare the results we obtain to other computations available in the literature, as a verification of our results and as a way to estimate the impact of missing corrections. We consider in this section the Higgs mass calculations in the Standard Model, and we will compare the results obtained with \SPheno with the computations performed at complete two-loop calculation in \cite{Buttazzo:2013uya}, and the full two-loop (plus leading three-loop) Higgs mass calculation implemented in the public code \texttt{SMH} \cite{Martin:2014cxa}.  These works take into account two-loop electroweak corrections, which are not available for generic theories and are not included in our code, hence we will quantify the size of these effects, together with effects from momentum, and investigate the discrepancy in masses coming from the different determination of the top Yukawa coupling. 

 It is interesting to examine the way that the two calculations avoid the Goldstone Boson Catastrophe. The calculation of Buttazzo et al.  \cite{Buttazzo:2013uya} was performed in Feynman gauge and using certain parameters on-shell, whereas the results implemented in {\tt SMH} are in a pure \MS scheme and Landau gauge, which is closer to our approach. In the latter paper, some resummation is performed by hand to eliminate the divergence in the mass calculation; it is perhaps surprising that  the absence of the function $\ov{V}(0,0,y,z)$ from the basis in {\tt TSIL} was not problematic, but there the calculation was performed by computing the set of integrals explicitly using {\tt TARCER} \cite{Mertig:1998vk} rather than starting from a set of generic expressions, so the result was found directly in terms of the other basis functions. In principle this should agree with our equation (\ref{EQ:Vtilde}).

For clarity, we recall that we define the (tree-level) Higgs potential as
\begin{equation}
 \label{EQ:pot_SM}
 \vtree=-\mu^2|H|^2+\lambda|H|^4, \qquad H = \twovec[G^+,\frac{1}{\sqrt{2}} (v+h + i G)].
\end{equation}

A first approach for the comparison between \SPheno and \cite{Buttazzo:2013uya} is to compute the Higgs mass with the quartic coupling $\lambda$ ranging in the interval $[0.125,0.130]$, and only setting the SM inputs to the same values as in \cite{Buttazzo:2013uya}, which we recall here
\begin{align}
\label{SMinputs_Buttazzoetal}
 G_F&=1.16638\times10^{-5}\text{ GeV}^{-2},\nn\\
 \alpha_s(M_Z)&=0.1184,\nn\\
 M_Z&=91.1876\text{ GeV},\nn\\
 m_t&=173.34\text{ GeV}.
\end{align}
They furthermore took the experimentally determined central value of the Higgs mass to be $125.15$~GeV, which we shall take as a reference value rather than an input. 
The use of consistent solutions to the tadpole equations -- as derived in \cite{Braathen:2016cqe} -- has also been implemented in the \SPheno code and this comparison in the context of the SM is a good occasion to study the effect of this additional shift to the tadpoles and mass diagrams, thus we compute the Higgs mass in this first method both with and without using the consistent tadpole solutions. 
A second approach to compute $m_h$ with \SPheno, which could potentially improve the comparison, is to use as well the same values for the top-Yukawa $y_t$ and electroweak gauge couplings $g_1,\,g_2$ as those given for each order in table 3 of \cite{Buttazzo:2013uya}. 

We obtain another result for $m_h$ with \texttt{SMH} \cite{Martin:2014cxa}, and although this code is made to perform Higgs mass calculations in the Standard Model to partial three-loop order, we use it here with the three-loop corrections always switched off, for the purpose of our comparison with \SPheno. 
We use the routine \texttt{calc\_Mh} that gives for a given loop order the value of $m_h$ from the inputs of the renormalisation scale $Q$, the quartic coupling $\lambda$, the top-Yukawa $y_t$, the Higgs VEV $v$, and the gauge couplings $g_3,\,g,\,g'$, all given at scale $Q$. 
In order to improve the comparison, we take the same values for the inputs as used at each order in \SPheno.
We give in table \ref{tab:compare_Buttazzoetal_mass} the values we find for the Higgs mass when taking the same values of $\lambda$ as found in \cite{Buttazzo:2013uya}, with the two methods described above for \SPheno and with \texttt{SMH}. 
\begin{table}[h]
\centering
\begin{tabular}{|c|c|c|c|c|c|}
\hline 
 & Value of & $m_h$ in $1^\text{st}$ approach & $m_h$ in $1^\text{st}$ approach & & \\
 \text{Loop order}& $\lambda$ found & without consistent & with consistent & $m_h$ in $2^\text{nd}$ & $m_h$ with \texttt{SMH}\\
  & in \cite{Buttazzo:2013uya} & tadpole solutions & tadpole solutions & approach & \\
\hline
 Tree level & 0.12917 & 125.79 GeV & 125.79 GeV & 125.79 GeV & 125.79 GeV\\
 One loop   & 0.12774 & 125.77 GeV & 125.77 GeV & 125.66 GeV & 126.10 GeV\\
 Two loops  & 0.12604 & 125.11 GeV & 125.08 GeV & 125.10 GeV & 125.46 GeV\\
\hline
\end{tabular}
\caption{Values of the Higgs mass at scale $Q=m_t$ for the values of the quartic couplings $\lambda$ found in \cite{Buttazzo:2013uya} at tree level, one loop and two loops, in the two approaches we used for \SPheno and with \texttt{SMH}. The first approach was to change only the SM parameter inputs while letting \SPheno determine the top-Yukawa and electroweak gauge couplings, and the Higgs mass is computed both with and without the consistent tadpole solutions. The second method was to take the same values of $y_t,\,g_1,\,g_2$ in \SPheno as in \cite{Buttazzo:2013uya} (and switch off the consistent tadpole routines). For \texttt{SMH}, the values of the input parameters -- the top-Yukawa, the electroweak gauge couplings, the Higgs VEV and the strong gauge coupling -- were taken from the outputs of the \SPheno scans. Computations are made with \SARAH-4.12.0, \SPheno-4.0.3 and \texttt{SMH}-1.0 \cite{Martin:2014cxa}.}
\label{tab:compare_Buttazzoetal_mass}
\end{table}

At tree-level, all the values we find with \SPheno and {\tt SMH} obviously match as the tree-level Higgs mass only depends on $\lambda$ and $v$ which have almost the same values here, and the divergence from the value of $125.15$ GeV is solely explained by the Higgs VEV which is not the same as in \cite{Buttazzo:2013uya} since they take it as an on-shell parameter, while we use the \MS value as described in \cite{Staub:2017jnp}. 
More importantly, the loop corrected values in the different methods also agree quite well, thanks to the improved determination of the top Yukawa coupling $y_t$ (including leading two-loop effects) recently implemented in \SARAH ~\cite{Staub:2017jnp}, and at each order in perturbation theory the Higgs masses we find are less than a GeV away from $125.15\text{ GeV}$. It is interesting to note that the values of $m_h$ found using the \SPheno code generated by \SARAH version 4.9.3 -- in which $y_t$ is only determined at one-loop order -- are approximately $2-2.5$ GeV below those shown in table \ref{tab:compare_Buttazzoetal_mass}, and hence illustrate the importance of the precise determination of the top Yukawa coupling for calculations of $m_h$. 
The small size of the difference between the values found with the couplings computed by \SPheno or taken from \cite{Buttazzo:2013uya} -- a few tens of MeV at two loops -- tend to indicate that the precision of the extraction of $y_t$ in \SPheno is now comparable to that in \cite{Buttazzo:2013uya}. Considering now the effect of the consistent tadpole solutions -- that appears only in the two-loop masses -- we observe a small shift of about 30 MeV to $m_h$, indicating that the perturbative expansion we perform in the tadpole equation is valid for the SM. 
Finally, the reasons explaining the remaining deviation of our results with respect to $125.15$ GeV are the following:
\begin{enumerate}
 \item[\textit{(i)}] the difference in the calculation of the Higgs VEV;
 \item[\textit{(ii)}] the two-loop electroweak corrections that are not (yet) implemented in \SARAH;
 \item[\textit{(iii)}] the momentum dependence currently missing at two loops in \SARAH.
\end{enumerate}
The different value of the Higgs VEV  is also quite certainly the main reason for the discrepancies between the values we obtain using \texttt{SMH} and those from \cite{Buttazzo:2013uya}. I.e. it is because we use the VEV computed in \SPheno in \smh, which does not correspond to the same accuracy of parameter extraction as used in \cite{Buttazzo:2013uya}, which would be required for a fair comparison directly between the two prior approaches: here our aim was to compare \emph{our} result separately with \cite{Buttazzo:2013uya} and \smh.

A further way to compare our results to those of \cite{Martin:2014cxa} and \cite{Buttazzo:2013uya} is to find for each order what value of the quartic Higgs coupling we need to obtain $m_h=125.15$ GeV, and our results are given in table \ref{tab:compare_Buttazzoetal_lambda}.
\begin{table}[h]
\centering
\begin{tabular}{|c|c|c|c|c|c|}
\hline 
  & & $\lambda$ in $1^\text{st}$ approach & $\lambda$ in $1^\text{st}$ approach  & & \\
 \text{Loop order}& $\lambda$ found in \cite{Buttazzo:2013uya} & without consistent &  with consistent & $\lambda$ in $2^\text{nd}$ & $\lambda$ with \texttt{SMH} \\
  & & tadpole solutions & tadpole solutions & approach & \\
\hline
 Tree level & 0.12917 & 0.12786 & 0.12786 & 0.12786 & 0.12786\\
 One loop   & 0.12774 & 0.12647 & 0.12647 & 0.12669 & 0.12580\\
 Two loops  & 0.12604 & 0.12613 & 0.12619 & 0.12614 & 0.12541 \\
\hline
\end{tabular}
\caption{Values of the Higgs quartic coupling $\lambda$ extracted from $m_h=125.15$ GeV, at tree level, one loop and two loops. The methods we used are explained in the caption of table \ref{tab:compare_Buttazzoetal_mass}. Values found using \SARAH-4.12.0 and \texttt{SMH}-1.0.}
\label{tab:compare_Buttazzoetal_lambda}
\end{table}
We observe that the change of $\lambda$ between each order of the perturbation expansion is approximately the same in all four methods. Moreover, the value we extract at two loops with \SPheno is very close to the value found in \cite{Buttazzo:2013uya}, only differing by $0.1\%$. 

\subsection{A detailed comparative study of \SPheno and {\tt SMH} results}
After this first comparison, we may now investigate in more depth the effects of the three sources of differences on the Higgs masses listed above, using \SPheno and \smh. To begin with, we should consider the Higgs VEV and its calculation: in \smh, calculations are performed in the Landau gauge, while \SPheno is by default set to use the Feynman gauge, and while the Higgs mass should in principle be gauge independent, its vacuum expectation value is not, hence there is an inconsistency coming from the use of a Feynman gauge VEV in \smh. The easiest way to correct this is to switch the \SPheno calculation to the Landau gauge -- we set in the code the gauge parameter $\xi$ to a very small finite value to approach the limit of the Landau gauge (the current implementation gives a numerical divergence when $\xi=0$) -- and then to use the new value of the Landau gauge VEV in \smh. The values we find for $m_h^{2\ell}$ with the two codes for the two different choices of gauge parameter and fixed values of $Q$ and $\lambda$ are given in  table \ref{tab:detailedcompare}. The first observation that can be made from these results is that the Higgs mass shows residual dependence on the gauge -- $m_h^{2\ell}$ varies by about 50 MeV between $\xi=1$ and $\xi=0.01$. 
This is explained mainly\footnote{In practice, there is always an additional residual gauge dependence as the Higgs mass is computed to finite order in perturbation theory and as not all parameters used for to compute $m_h$ are determined to the same loop order. } by the difference in the calculation of the \MS value for the electroweak VEV in \SPheno between the Feynman gauge and other gauges: in the case of Feynman gauge one loop corrections from $\delta_{VB}$ as well as two-loop corrections from $\delta_r$ are included which are not available in general $\xi$ (see the appendix A of \cite{Staub:2017jnp} for details of the matching in Feynman gauge). On the other hand, in $\xi$ gauge, the VEV is calculated from $M_Z^{2,\text{\MS}} = 1/4 (g_1^2+g_2^2) v^2 = M_Z^{2,\text{pole}} - \Pi_{ZZ}^T$ where $\Pi_{ZZ}^T$ is the transversal self-energy of the Z-boson at one-loop.
What is more interesting is that the agreement between the two codes improves greatly once we use the Landau gauge in \SPheno; indeed the difference in the Higgs mass results is reduced from approximately 0.4 GeV to less than 0.05 GeV.

A second point we can study is the effect of the two-loop momentum dependence and two-loop electroweak corrections. Let us introduce the notation for calculating the pole mass via
\begin{align}
m_h^2 =& \,\,2 \lambda v^2 + \Delta^{(1)} M_h^2 (m_h^2) + \Delta^{(2)} M_h^2 (m_h^2)
\end{align}
where
\begin{align}
\Delta^{(\ell)} M_h^2 (s) \equiv& - \frac{1}{v}\frac{\partial \Delta V^{(\ell)}}{\partial h}\bigg|_{h=G=G^+=0} +\Pi_{hh}^{(\ell)} (s) \nn\\
\equiv& \,\,\mathrm{div} \bigg[ \Pi_{hh}^{(\ell)} (s) \bigg] + \ov{\Delta}^{(\ell)} M_h^2 (0) +\mathcal{O} (s),
\end{align}
where $\mathrm{div} \bigg[ f(s) \bigg] $ denotes all terms in $f(s)$ that diverge as $s\rightarrow 0$.
Our \SPheno code computes the one-loop corrections in any $R_\xi$ gauge with full momentum dependence, but the two-loop corrections are performed in a generalised effective potential approach -- $i.e.$ we keep only the divergent part of the momentum dependence (see section 4 of \cite{Braathen:2016cqe} for more details). The momentum in the two-loop routines is fixed (for speed of calculation) whereas that in the one-loop routines is adjusted to solve the on-shell condition:
\begin{align}
s =& \,\,2\lambda v^2 + \Delta^{(1)} M_h^2 (s) + \Delta^{(2)} M_{h,\SPheno}^2 (s), \nn\\
\Delta^{(2)} M_{h,\SPheno}^2 (s) \equiv& \,\,\mathrm{div} \bigg[\Pi_{hh, \mathrm{gaugeless}}^{(2)} (s)\bigg] +\Delta^{(2)} M_{h, \mathrm{gaugeless}}^2(0).
\end{align}
This begs the question of how to compare our result with \smh: ideally, we would like to extract a result from \smh  which is comparable to ours. However, this is confounded by several factors:
\begin{enumerate}
\item It is impossible to extract the electroweak contributions in \smh, because the result is not finite as the electroweak gauge couplings become zero. 
\item To avoid the Goldstone boson catastrophe and ensure cancellation between Goldstone boson and longitudinal gauge boson diagrams, in the two-loop corrections in \smh the external momentum $s$ has been replaced by $2\lambda v^2$ wherever it appears in a pre-factor (but not in the arguments of the loop functions). 
\item The term
\begin{equation}
\label{eq:missingterm}
\Delta^{(1)} M_h^2 \supset \frac{3 \lambda}{16\pi^2} (s^2 - 4 \lambda^2 v^4) \frac{B(0,0)}{2\lambda v^2} ,
\end{equation}
which is part of the one-loop correction coming from Goldstone bosons and longitudinal gauge bosons, is moved into the two-loop corrections, with the justification that on-shell $s = 2 \lambda v^2 + \Delta^{(1)} M_h^2 $ so will give a contribution at two-loop order when solving for the on-shell mass. 
\end{enumerate}

If it were not for point (2) above, it would perhaps have been possible to extract the result for the generalised effective potential approximation for the electroweak corrections. Instead, we will simply compare the results as we vary the momentum in \smh; by modifying slightly the source code, we obtain a version of \smh without the momentum dependence at two loops (but retaining the dependence at one loop). Interestingly, the result of \smh is finite even when $s=0$ meaning that the divergence as $s\rightarrow 0$ has been removed. It turns out that this is because of the term (\ref{eq:missingterm}), which has the effect of cancelling the divergences as $s\rightarrow 0$ (even though this cancellation is fictitious). If we write $\delta^{(2)}(s)$ for the missing momentum dependence in \smh from setting the coefficients of loop functions equal to $2\lambda v^2$, then we have
\begin{align}
\Delta^{(2)} M_{h,\smh}^2 (s) =&\frac{6 \lambda}{16\pi^2} \left(\Delta^{(1)} M_{h}^2 (s)\right) B(0,0) +  \Delta^{(2)} M_{h, \mathrm{gaugeless}}^2 (s) + \Delta^{(2)} M^2_{h, \mathrm{electroweak}} (s) + \delta^{(2)} (s)\nn\\
=& \frac{6 \lambda}{16\pi^2} \left(-\frac{6 \lambda^2 v^2}{16\pi^2} B(0,0) + \Delta^{(1)} M_{h,{\tt SMH}}^2 (0)\right) B(0,0)\nn\\
& + \mathrm{div} \bigg[\Pi_{hh, \mathrm{gaugeless}}^{(2)} (s) + \Pi_{hh, \mathrm{electroweak}}^{(2)} (s) + \delta^{(2)} (s)\bigg]\nn\\
&  +  \ov{\Delta}^{(2)} M_{h, \mathrm{gaugeless}}^2(0) + \ov{\Delta}^{(2)} M^2_{h, \mathrm{electroweak}}  (0) + \bar{\delta}^{(2)} (0)  + \mathcal{O} (s),\nn\\
=& \frac{12 \lambda}{16\pi^2} \left(\ov{\Delta}^{(1)} M_{h}^2 (0)\right) +  \Delta^{(2)} M_{h, \mathrm{gaugeless}}^2(0) + \Delta^{(2)} M^2_{h, \mathrm{electroweak}}  (0) + \bar{\delta}^{(2)} (0) + \mathcal{O} (s).\nn
\end{align}
The cancellations of the divergences imply that 
\begin{align}
\mathrm{div} \bigg[\Pi_{hh, \mathrm{gaugeless}}^{(2)} (s) + \Pi_{hh, \mathrm{electroweak}}^{(2)} (s) + \delta^{(2)} (s)\bigg] \overset{?}{=}& \frac{1}{(16\pi^2)^2} \bigg[ 36\lambda^2v^2 \blog^2 (-s) - 72\lambda^2 v^2\blog(-s)\bigg]  \nn\\
&+ \frac{6\lambda}{16\pi^2}\left(\ov{\Delta}^{(1)} M_{h}^2 (0)\right) \blog(-s),
\end{align}
where 
\begin{align}
(16\pi^2)\ov{\Delta}^{(1)} M_{h}^2 (0) \equiv& (16\pi^2)\Delta^{(1)} M_{h,{\tt SMH}}^2 (0) - 12\lambda^2 v^2 \nn\\
=&  - 12 \lambda^2 v^2 + 18 \lambda^2 v^2 \blog (m_h^2) - 12 y_t^2 m_t^2 \blog (m_t^2) \nn\\
& + \left(\frac{g_Y^2 + g_2^2}{2}\right) m_Z^2 \bigg[ 3 \blog m_Z^2  + 2 \bigg] + g_2^2 m_W^2\bigg[ 3 \blog m_W^2  + 2 \bigg].
\end{align}

On the other hand, by evaluating the diagrams for the Standard Model in the gaugeless limit retaining only the top Yukawa coupling and the Higgs quartic $\lambda$ we find
\begin{align}
 \mathrm{div} \bigg[\Pi_{hh, \mathrm{gaugeless}}^{(2)} (s)\bigg] = \frac{6\lambda v^2}{(16\pi^2)^2}  \blog (-s) \bigg[&\lambda^2 \bigg( -14 + 18 \blog (m_h^2) + 3 \blog (-s) \bigg)\nn\\
& - 2 y_t^2 \bigg( \lambda + ( y_t^2 - \lambda) \blog (m_t^2) \bigg) \bigg],
\end{align}
so we can see there are several remaining pieces that must be cancelled by \\$\mathrm{div} \bigg[ \Pi_{hh, \mathrm{electroweak}}^{(2)} (s) + \delta^{(2)} (s)\bigg] $.
 But if we set $s^2_{\rm fixed} = -Q^2 $ in our routines we should cancel the divergent part exactly, and leave us only with $\Pi_{hh, \mathrm{gaugeless}}^{(2)} (0). $ We can then determine
\begin{align}
 \ov{\Delta}^{(2)} M^2_{h, \mathrm{electroweak}} (0) + \ov{\delta}^{(2)} (0) =& \Delta^{(2)} M_{h,\smh}^2  (0) - \Delta^{(2)} M_{h,\SPheno}^2 (-Q^2) - \frac{12 \lambda}{16\pi^2} \left(\ov{\Delta}^{(1)} M_{h}^2 (0)\right).
\end{align} 
We find that this residual difference is tiny; at $Q=m_t=173.34\text{ GeV}$ with $\lambda = 0.12604, y_t=0.9345, v=247.07$ GeV and the gauge couplings $(g_3, g_2, g_Y) = (1.1654, 0.6442, 0.2782)$ we have: 
\begin{align}
 \ov{\Delta}^{(2)} M^2_{h, \mathrm{electroweak}} (0) + \ov{\delta}^{(2)} (0) \simeq -0.03 (\mathrm{GeV})^2 = -0.0002\% m_h^2!
\end{align}
This corresponds to a tiny value of the electroweak corrections; a similar observation was made in \cite{Buttazzo:2013uya}. 

\begin{figure}[h]
 \centering
 \includegraphics[width=.8\textwidth]{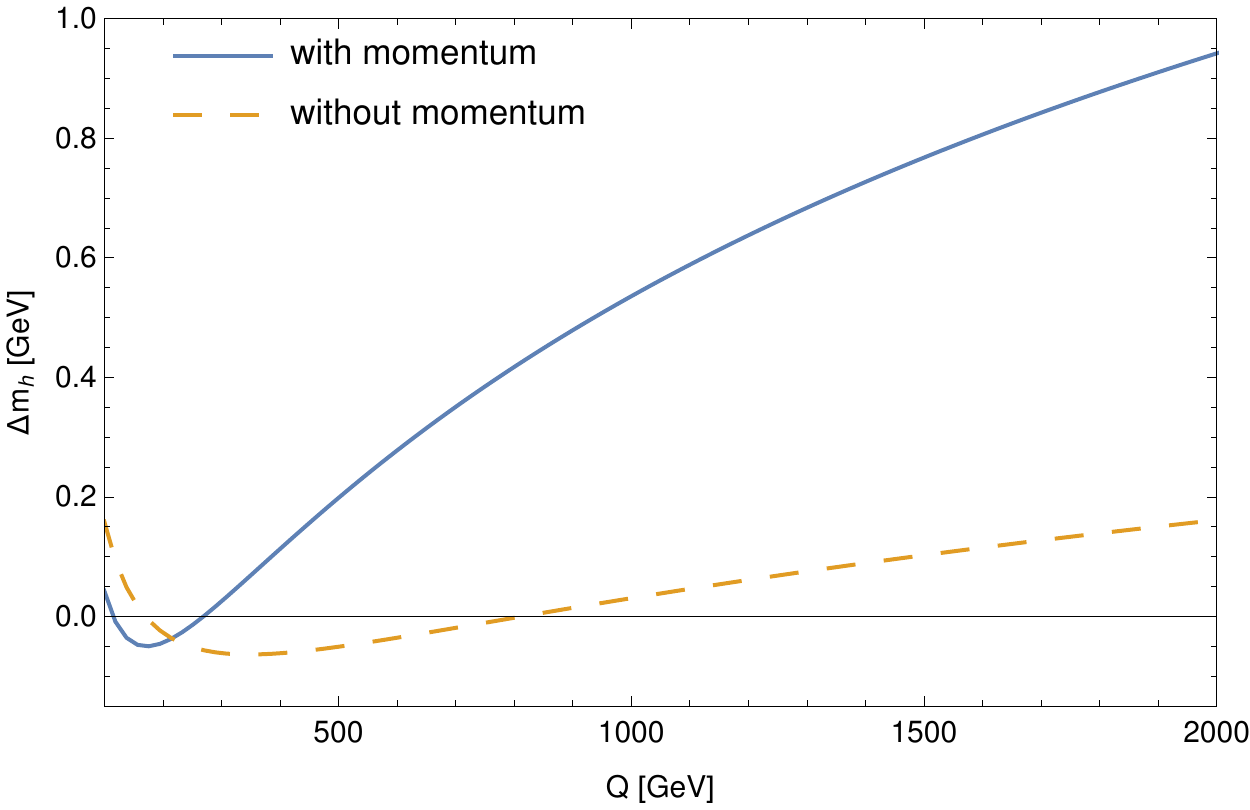}
 \caption{Difference between the two-loop Higgs mass computed by \smh and \SPheno -- $(m_h^{2\ell})^\SPheno-(m_h^{2\ell})^\smh$ -- as a function of the renormalisation scale $Q$, with (\textit{blue curve}) and without (\textit{orange dashed curve}) the momentum dependence at two loops in \smh. The Higgs quartic coupling is here $\lambda=0.12604$. In \SPheno the contributions of the light SM fermions are turned off and the external momentum in the two-loop routines is set to $s=m_h^2|^\text{tree}$.}
 \label{fig:compSPhenoSMH}
\end{figure}

Finally, we compare the more physically meaningful differences between the codes when we take $s=m_h^2|^\text{tree}$ in our routines. The values of the Higgs mass computed with \SPheno after turning off the light SM fermion contributions and with the modified version of \smh is given in table \ref{tab:detailedcompare}, and strikingly they only differ by $40$ MeV when we include the momentum dependence in \smh$\,$-- in other words, for $Q=173.34$ GeV, the momentum dependence and electroweak corrections amount to only 0.03\% of $m_h$. We further examine the importance of both the momentum dependence and EW corrections by varying now the renormalisation scale at which we compute the Higgs mass: for this purpose, figure \ref{fig:compSPhenoSMH} shows the difference of the two-loop masses between the two codes -- more precisely $(m_h^{2\ell})^\SPheno-(m_h^{2\ell})^\smh$ -- with and without momentum, as a function of the renormalisation scale $Q$ (where the \MS parameters are extracted by \SPheno at each value while keeping $\lambda$ fixed rather than evolving the parameters: the idea is to show the importance of the choice of scale rather than the stability of the computation). While for large scales the two-loop momentum effects may become large (1 GeV or more), the electroweak corrections represent at most 0.2 GeV and even vanish for a scale close to the \MS top mass.

\begin{table}[ht]
\centering
\begin{tabular}{|c||c|c|c||c|c||}
\hline 
 & \multicolumn{3}{|c||}{\SPheno} & \multicolumn{2}{c||}{\smh} \\
\hline
 $\xi$  & 1 & 0.01 & 0.01 & \multicolumn{2}{|c||}{0} \\
\hline
$v$ (GeV) & 247.494 & \multicolumn{4}{c||}{ 246.914} \\\hline
$y_t$ & 0.939 & 0.939 & \multicolumn{3}{c||}{0.940} \\\hline
$(g_3,g_2,g_Y)$ & \multicolumn{5}{|c||}{$(1.1654, 0.6452, 0.2780)$} \\ \hline
\hline 
 $2\ell$ momentum & partial & partial & partial &  full & none  \\
 dependence & $s=m_h^2|^\text{tree}$ & $s=m_h^2|^\text{tree}$ & $s=m_h^2|^\text{tree}$ &  iterative & $s=0$ \\
\hline
 Light SM fermions & yes & yes & no &  no & no \\
\hline
 $m_h^{2\ell}$ (GeV) & 125.083 & 125.134 & 125.133 & 125.176 & 125.121 \\ 
\hline
\end{tabular}
\caption{Comparison of two-loop Higgs masses calculated with the codes \SPheno and \smh, for different choices of gauge in \SPheno and switching on and off the two-loop momentum dependence in \smh. The renormalisation scale is fixed to $Q=173.34$ GeV, and the Higgs quartic coupling is $\lambda=0.12604$ and is not varied (the idea being to illustrate the importance of the choice of scale, rather than the stability of the result). All other inputs for \smh are taken to the same values as in \SPheno. In \SPheno the only two-loop momentum dependence is from pseudo-scalar diagrams and only a generalised effective potential approach (see main text) with $s=m_h^2|^\text{tree}$, while in \smh the full two-loop dependence is implemented and is used to find $m_h$ iteratively. }
\label{tab:detailedcompare}
\end{table}

\subsection{Momentum dependence}
Implementing the solution to the Goldstone boson catastrophe in \SARAH has required the insertion of external momentum in infra-red divergent loop integrals, and thus we should also investigate the impact of the momentum $s=-p^2$ on the Higgs mass calculation in \SPheno. In practise, we have set for the majority of scans the external momentum for the two-loop calculations to be equal to $m_h^2|^\text{tree}$ but we will now vary the momentum to study its impact on $m_h$. Table \ref{table:SM_mh_p2} shows the shift to the two-loop Higgs mass -- with respect to the value computed with $s=(125 \text{ GeV})^2$ -- for external momentum in loops equal to $s=\alpha\times(125 \text{ GeV})^2$, where $\alpha$ ranges from $10^{-6}$ to $10^6$ and for $\lambda=0.126$ and $\lambda=0.130$. For all values of the external momentum considered here, the variation of the Higgs mass remains small: at most they become of order $\sim0.13x1$ GeV for $\alpha=10^{-6}$ ($i.e.$ $\sqrt{s}=0.125$ GeV), and while this effect is noticeable, it is far from the divergences that could have been feared when approaching the limit of $s\rightarrow 0$. All in all, although pole masses -- as we compute here -- are in principle found as the zero of the inverse propagator, that has to be found iteratively as the self-energy contains momentum dependence, we see from the minute effects of momentum in the range $\alpha\in[1/2,100]$, relevant for scalar masses, that we will not require an iterative solution and that simply taking $s=(125\text{ GeV})^2$ in the loop diagrams with pseudo-scalars will be a satisfactory approximation. In particular, changing $s$ between $m_h^2|^\text{tree}$ and 125 GeV causes a difference in $m_h^{2\ell}$ of less than an MeV. 

\begin{table}[ht]
 \centering
 \begin{tabular}{|c|c|c|c|c|c|c|c|c|c|c|}
  \hline
  Value of & \multicolumn{9}{c|}{Shift to the two-loop Higgs mass for the values of the momentum $s=\alpha \times(125\text{ GeV})^2$, in GeV}\\
  \cline{2-10}
  $\lambda$ & $\alpha=10^{-6}$ & $\alpha=10^{-4}$ & $\alpha=10^{-2}$ & $\alpha=1/2$ & $\alpha=1$ & $\alpha=2$ & $\alpha=100$ & $\alpha=10^4$ & $\alpha=10^6$ \\ 
  \hline
  0.126 & 0.1210 & 0.0655 & 0.0252 & 0.0028 & 0.0 & -0.0025 & -0.0100 & -0.0048 & 0.0155 \\
  0.130 & 0.1302 & 0.0704 & 0.0270 & 0.0030 & 0.0 & -0.0026 & -0.0106 & -0.0048 & 0.0560 \\
  \hline
 \end{tabular}
\caption{Shift in GeV of the two-loop Higgs mass in the Standard Model -- computed with \SPheno and with respect to the value obtained for $p=125$ GeV -- for different values of the quartic coupling $\lambda$, and of the incoming momentum $s$ in the two loop routines.}
\label{table:SM_mh_p2}
\end{table}

We emphasise however that the effect of momentum on Goldstone boson mass diagrams discussed here is only a subset of the general momentum dependence of the two-loop masses, which should in principle be taken into account, as seen in the previous subsections.

\section{The NMSSM}
\label{sec:SUSY}

As a second check of our new solution, and demonstration of its importance, we shall compare the results for the three different options to solve the Goldstone Boson Catastrophe in the example of the Next-to-minimal supersymmetric standard model (NMSSM) -- see \cite{Ellwanger:2009dp} and references therein for a detailed description of the model. Indeed, the NMSSM is the first supersymmetric model for which the problems at certain points in the parameter space were found in earlier versions of \SARAH. Here we shall show that this is avoided, and have a preliminary look at the impact of the ``consistent tadpole solutions.''

We start with a test point defined by
the following input parameters\footnote{Note, in this section we use $\lambda$, which is in all other sections the quartic Higgs coupling from the SM, for the superpotential coupling $\hat S \, \hat H_d\, \hat H_u$ as usually done in the NMSSM.}:
\begin{eqnarray}
\label{eq:NMSSM_p1}
& \lambda=0.7,\quad \kappa=0.25,\quad A_\lambda=1350~\text{GeV},\quad A_\kappa=-500~\text{GeV},\quad \mu_{\rm eff} = 600~\text{GeV} &  \\
\nonumber 
& M_1=M_2=1000~\text{GeV},\quad  M_3 = 2000~\text{GeV},\quad T_{u,33} = 1500~\text{GeV},\quad m_{\tilde u,33} = 1000~\text{GeV} &
\end{eqnarray}
and all other soft-masses set to 2~TeV. The Higgs masses for the following calculations are given in Table~\ref{tab:NMSSMp1}:
\begin{enumerate}
 \item $D$-terms turned off in mass matrices but retained in tadpole solutions (as in previous versions of \SARAH), labelled ``$D$'' in the table.
 \item Regulator masses with $R=10^{-5}$--$10^{-1}$.
 \item Goldstones set on shell, with and without consistent tadpole solutions, labelled OS and OS$+$Tad respectively.
\end{enumerate}
We see from this table that there is an agreement in the light Higgs mass of about 0.4~GeV between all the calculations  if $R$ is chosen to be about $10^{-2}$.  
\begin{table}[tb]
\begin{tabular}{|c|c|c c c c c |c c|}
\hline
      &   $D$   & $R=10^{-5}$ & $R=10^{-4}$ & $R=10^{-3}$ & $R=10^{-2}$ & $R=10^{-1}$ & OS & OS+Tad \\
\hline 
$h_1$ & 129.58   & 65.27    & 124.63   & 129.07    & 129.82   & 130.58   & 129.70   &129.97   \\
$h_2$ & 315.64   & 312.84   & 315.39   & 315.59    & 315.55   &  315.67  & 315.09   &315.60   \\
$h_3$ & 1632.28  & 1627.55  & 1631.77  & 1632.36   & 1632.63  & 1632.81  &  1632.51 &1633.39  \\
$A_1$ &  582.02  & 582.61   &  582.31  &  582.02   &  581.74  & 581.63   &  580.94  &581.23   \\
$A_2$ & 1631.98  & 1630.38  &  1631.15 &  1631.88  &  1632.43 & 1632.59  & 1632.04  & 1632.60 \\
\hline
\end{tabular}   
\caption{The Higgs masses in the NMSSM (in GeV) for the parameter point defined by eq.~(\ref{eq:NMSSM_p1}) for different choices for the two-loop corrections. }
\label{tab:NMSSMp1}
\end{table}

While the new ``on-shell'' solution of the Goldstone boson catastrophe is optimal, between introducing a regulator $R$ and the previous approach with neglected $D$-terms in the scalar mass matrix, the latter is preferred because one does not need to check for a suitable choice of $R$ to stabilise the results. However, we can now consider parameter points where the old method fails. 
\begin{figure}[tb]
\centering
\includegraphics[width=0.5\linewidth]{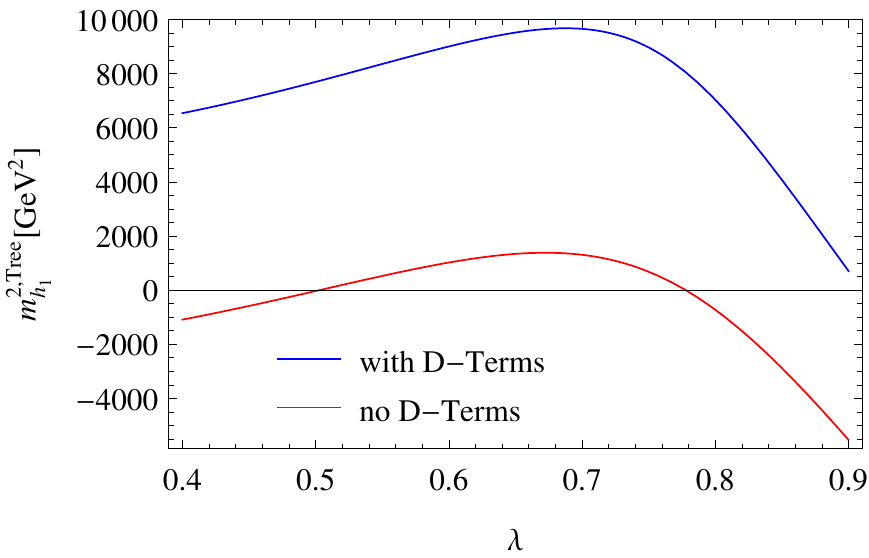} 
\caption{The lightest scalar mass squared for the parameter point defined by eq.~(\ref{eq:NMSSM_p2}) when calculating with and without $D$-term contributions.}
\label{fig:NMSSM_mh1}
\end{figure}
This is shown for the point defined by 
\begin{eqnarray}
\label{eq:NMSSM_p2}
&  \kappa=0.6,\quad A_\lambda=200~\text{GeV},\quad A_\kappa=-200~\text{GeV},\quad \mu_{\rm eff} = 150~\text{GeV} &  \\
\nonumber 
& M_1=M_2=1000~\text{GeV},\quad  M_3 = 2000~\text{GeV}&
\end{eqnarray}
and all scalar soft-masses set to 2~TeV. The lightest scalar tree-level mass with and without the $D$-terms as function of $\lambda$ is shown in figure~\ref{fig:NMSSM_mh1}.
One can see that for $\lambda \simeq 0.5,0.8$, the lightest scalar becomes massless in the limit of vanishing $D$-terms. Thus, for these values, divergences in the two-loop corrections can be expected which are this time not associated with the Goldstone but with the lightest CP even state. 
\begin{figure}[tb]
\centering
\includegraphics[width=0.67\linewidth]{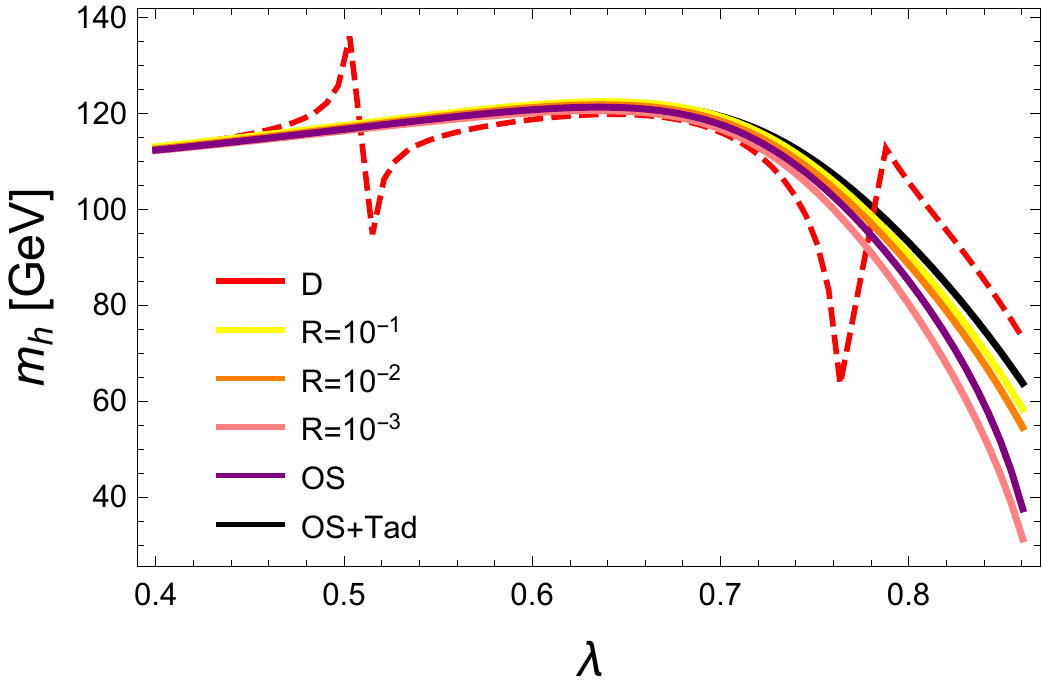}
\caption{The lightest Higgs mass at the two-loop level for the parameter point defined by eq.~(\ref{eq:NMSSM_p2}) for different methods to regulate the two-loop corrections. }
\label{fig:NMSSMp2}
\end{figure}
We show the lightest Higgs mass in figure~\ref{fig:NMSSMp2} as function of $\lambda$ for different methods to regulate the two-loop corrections. Obviously, the approach of neglecting electroweak $D$-terms fails for values of $\lambda$ at which the masses entering the loop calculations become very light. However, for very large values of $\lambda$ which are away from the poles, the agreement with the other calculations is also rather poor. In contrast, over the entire range of $\lambda$ we see a good agreement between the methods using regulator masses, if $R=10^{-2}$ or $10^{-3}$ is chosen, and the method of treating the Goldstones on-shell. It is interesting that for these values of $R$ the minimum mass is $\sqrt{R}\times M_{\rm SUSY} \simeq 100$ GeV, i.e. logarithmic contributions involving the light scalars are being excised. 

We note that the corrections from the consistent tadpole solution are small until $\lambda$ becomes large, at which point we see significant deviations. However, as $\lambda$ approaches $0.9$ we see from figure \ref{fig:NMSSM_mh1} that the tree-level lightest Higgs mass approaches zero, so we expect our perturbative calculation of the ``consistent tadpole solution'' to break down and become unreliable.

\section{Split SUSY}
\label{SEC:SPLIT}

In Split SUSY scenarios \cite{ArkaniHamed:2004fb, Giudice:2004tc, ArkaniHamed:2004yi,Kilian:2004uj,Bernal:2007uv, Giudice:2011cg}, the SUSY scalars are much heavier than the gauginos and Higgsinos. Consequently, these models should be studied in an effective approach where all SUSY scalars
are integrated out at some matching scale. The Lagrangian below this scale is given by
\begin{align}
\lagr = & \lagr_{\rm SM} - 
\left(\frac12 M_3 \tilde{g}^\alpha \tilde{g}^\alpha + \frac12 M_2 \tilde{W}^a \tilde{W}^a + \frac12 M_B \tilde{B}\tilde{B} + \mu \tilde{H}_u^T \epsilon \tilde{H}_d \, + \, \text{h.c.}  \right) \\
& -\left[\frac{1}{\sqrt{2}} H^\dagger \left(\tilde{g}_{2u}  \sigma^a \tilde{W}^a + \tilde{g}_{1u} \tilde{B} \right) \tilde{H}_u +   
\frac{1}{\sqrt{2}} H^T \epsilon \left(-\tilde{g}_{2d}  \sigma^a \tilde{W}^a + \tilde{g}_{1d} \tilde{B} \right) \tilde{H}_d \, + \, \text{h.c.} \right]
\label{EQ:LagSplit}\end{align}
where $\lagr_{\rm SM}$ is the Standard Model Lagrangian with Higgs potential (\ref{EQ:pot_SM}). Because of the matching between the effective, non-supersymmetric model and the MSSM, the quartic Higgs coupling $\lambda$ 
as well as the new Yukawa-like interactions $\tilde{g}_{(1,2)(u,d)}$ are not free parameters but fixed by the matching conditions at the scale $M_M$. At tree-level, the following relations hold
\begin{align}
\tilde{g}_{2u}(M_M) = & g_2(M_M) \sin\beta \\
\tilde{g}_{2d}(M_M) = & g_2(M_M) \cos\beta \\
\tilde{g}_{1u}(M_M) = & \sqrt{\frac35} g_1(M_M) \sin\beta \\
\tilde{g}_{1d}(M_M) = & \sqrt{\frac35} g_1(M_M) \cos\beta \\
\lambda(M_M) =& \frac{1}{8} (g_1^2(M_M) + g_2^2(M_M)) \cos^22\beta
\end{align}
Here, $g_1$ and $g_2$ are the running gauge couplings of $U(1)_Y \times SU(2)_L$ and $\beta$ is defined as the mixing angle of the two Higgs doublets in the MSSM (in contrast to the definition in the MSSM as a ratio of expectation values). There are important higher order corrections to the matching conditions which are necessary to have a precise prediction for the Higgs mass at the 
low scale. In particular $\lambda$ has been calculated including the two-loop SUSY  corrections \cite{Bagnaschi:2014rsa,Vega:2015fna,Bagnaschi:2017xid}. The numerical value of these corrections depends on the many SUSY parameters at the matching scale; however,  a commonly taken useful approximation is to give the scalars a common mass $\Mm$, in which case the corrections can be given in terms of just this scale and the squark mixing. Moreover, in strict split SUSY where the fermion masses are protected by an R-symmetry (or another symmetry in Fake Split SUSY \cite{Benakli:2013msa,Benakli:2015ioa}) near the electroweak or TeV scale and well below $\Mm$, the squark mixing must by very small. In which case, the leading corrections to the Higgs quartic coupling are purely electroweak at one loop, and at two loops contain no logarithmic terms -- meaning that they are very small (in particular since the strong gauge and top Yukawa couplings run to small values at higher scales), so using the tree-level relationship above can be good enough. 

Below the scale $\Mm$, we must run to the scale of the fermion masses, before also integrating them out, and then running to the electroweak scale in the Standard Model. In some previous approaches, the running was performed all the way down to the electroweak scale, before calculating the Higgs mass in the full Lagrangian (\ref{EQ:LagSplit}); however, it was found in \cite{Benakli:2013msa} that in this approach it is necessary to include the \emph{three loop} leading logarithm involving the gluino mass to obtain good agreement between the two results -- this is  automatically resummed by the renormalisation group running in the former approach. In either case, the full contribution of the gauginos and Higgsinos to the matching conditions is only known in the literature to \emph{one loop} order\cite{Bagnaschi:2014rsa}. 

\begin{figure}[tb]
\centering
\includegraphics[width=0.4\linewidth]{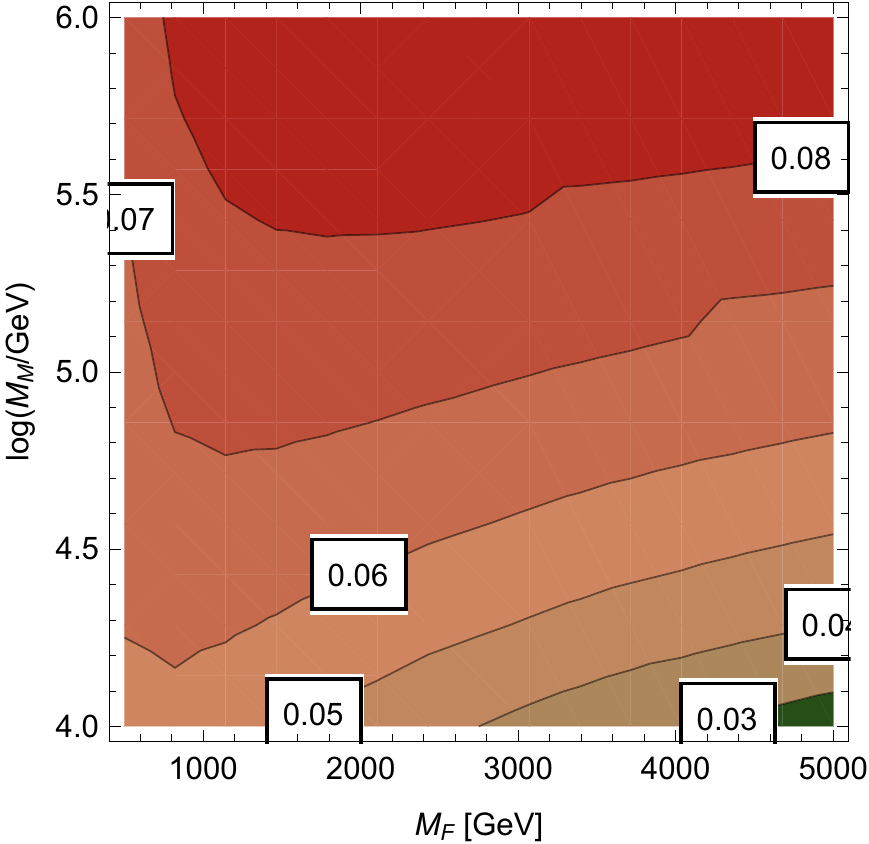} \quad
\includegraphics[width=0.4\linewidth]{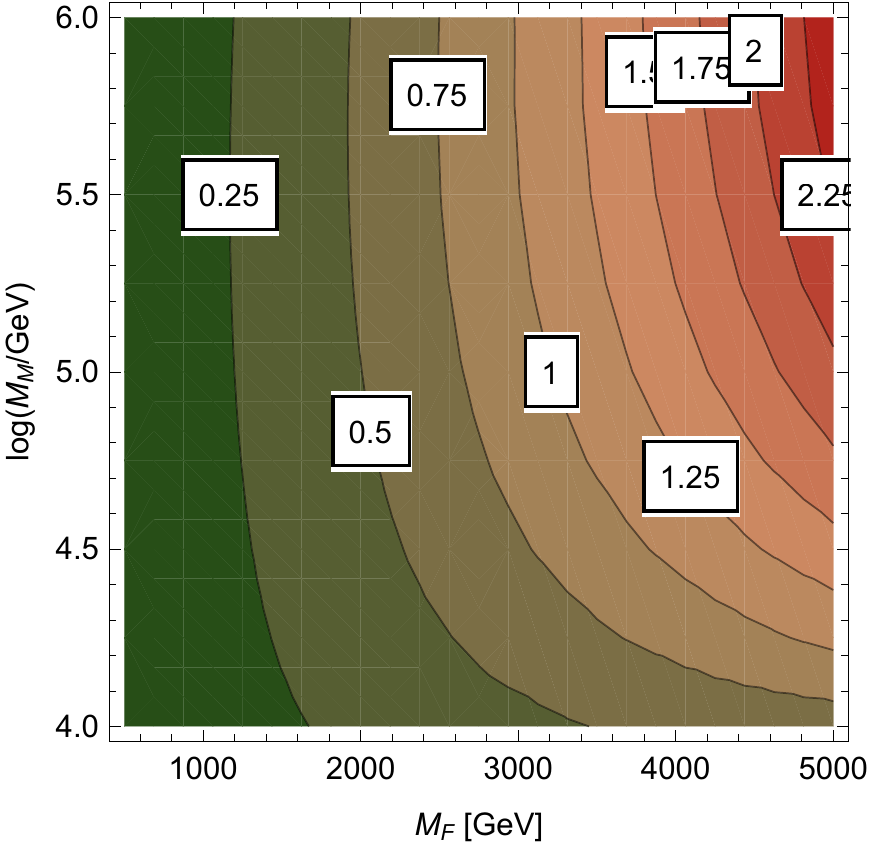} \\
\includegraphics[width=0.4\linewidth]{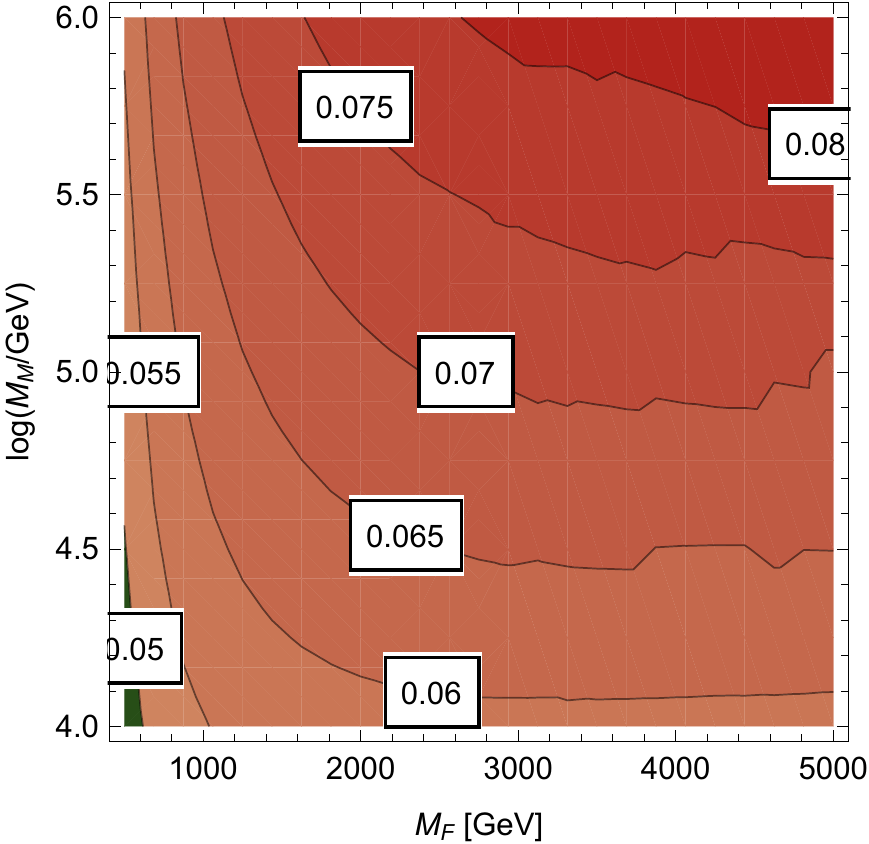} \quad
\includegraphics[width=0.4\linewidth]{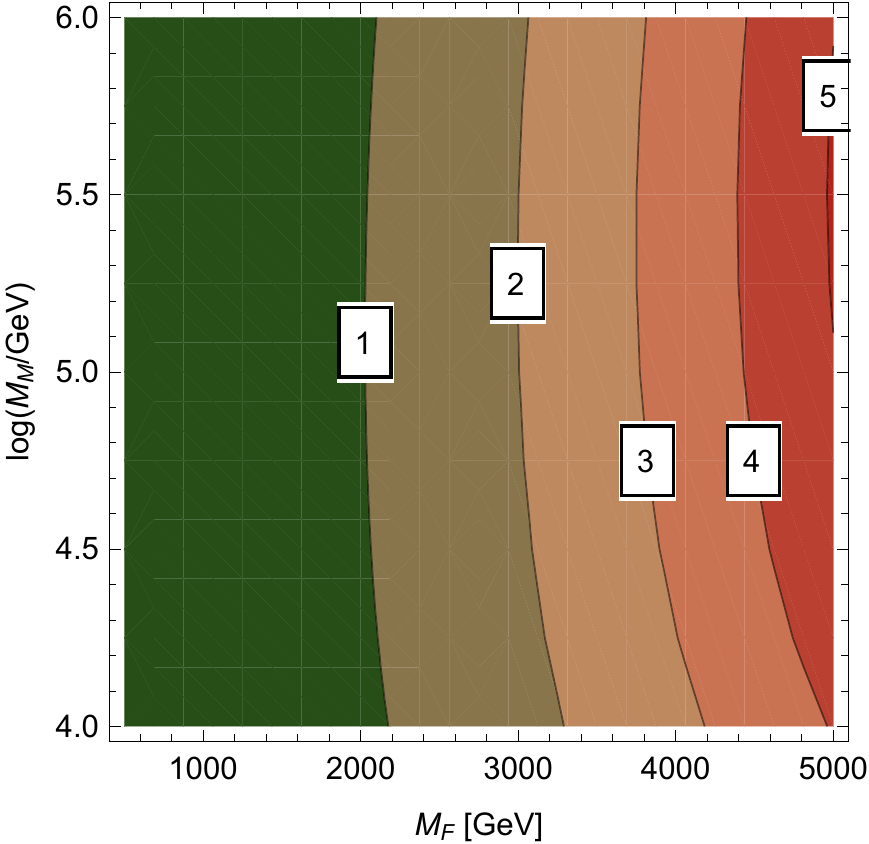}
\caption{The change in the Higgs mass in GeV due to the two-loop corrections involving the new Yukawa-like interactions $\tilde{g}_{(1,2)(u,d)}$. On the top, we used $\tan\beta=1$ at $M_M$ and on 
the bottom $\tan\beta=10$. The left plots are with the consistent tadpole solutions, the right ones without.}
\label{fig:SplitSUSY}
\end{figure}

Hence in this section we are interested in the effect of the two-loop corrections to the Higgs mass stemming from the $\tilde{g}_{(1,2)(u,d)}$ couplings which have not been studied in the literature before. They are expected to be small since they originate from electroweak interactions at the matching scale (and so, admittedly, one could argue that we should neglect them in the gaugeless limit). We shall not discuss the absolute value of the Higgs mass, for which we would need to include all higher order corrections to the matching that have been calculated elsewhere, but only on the impact of the new two-loop corrections. The overall size of these corrections is rather insensitive to the exact matching conditions and we are using the above tree-level relations; but as we noted earlier, these should be a particularly good approximation for larger matching scales. 
 
We make in addition the simplifying assumption that at $M_M$ the SUSY fermions are degenerate, i.e. 
\begin{align}
\mu(M_M) = M_1(M_M) = M_2 (M_M)= M_3 (M_M)\equiv M_F 
\end{align}
and thus we are left with three free parameters:
\begin{eqnarray*}
M_F \,,\quad M_M\,,\quad \tan\beta. 
\end{eqnarray*}
\SARAH uses two-loop RGEs for the running between $M_M$ and the renormalisation scale $Q$ which we set to $M_F$, that as mentioned above is necessary to avoid large logarithmic contributions from the gluino. 
The size of the two-loop corrections proportional to the $\tilde{g}_{(1,2)(u,d)}$ couplings in the $(M_M,M_F)$ plane is shown in figure~\ref{fig:SplitSUSY} for $\tan\beta=1$ and 10 for a calculation with and without the consistent tadpole solutions explained in sec.~\ref{SEC:Consistent}. 
We show here results for $M_F$ up to 5 TeV. In order not to increase the theoretical uncertainty in the presence of new fermions in the multi-TeV range, we made use of the functionality in \SARAH to perform the Higgs mass calculation in the effective SM \cite{Staub:2017jnp}. For this purpose, a second matching is performed to extract $\lambda$ at the renormalisation scale $Q$. The imposed matching condition is 
\begin{equation}
m_h^{\rm SM}(M_F) \equiv m_h^{\rm Split}(M_F) 
\end{equation}
i.e. we perform a matching of the Higgs pole masses as suggested in \cite{Athron:2016fuq}, from which an effective $\lambda$ is derived. $\lambda$ is then evolved to $m_t$ using three-loop RGEs of the SM. At $m_t$ the Higgs mass is calculated within the SM at the two-loop level. The additional loop-corrections discussed here enter the calculation of $m_h^{\rm Split}(M_F)$, i.e. the calculation of $\lambda^{\rm SM}(M_F)$. \\
We see that the additional corrections for SUSY fermions are always well below 1~GeV once the consistent solution to the tadpole equations are included. However, if those are not used, the misleading impression of sizeable corrections of a few GeV is given; it would be interesting to investigate this phenomenon further.

\section{Two-Higgs-Doublet Model}
\label{SEC:THDM}

The scalar potential of the CP-conserving 2HDM is defined in terms of scalar $SU(2)_L$ doublets in a basis $\{\Phi_1,\Phi_2\}$ -- sometimes called the $\mathbb{Z}_2$ basis -- as  
\begin{align}
\label{pot_Z2_basis}
  \vtree=&\,\,m_{11}^2 \left(\Phi_1^\dagger\cdot \Phi_1\right) + m_{22}^2 \left(\Phi_2^\dagger\cdot \Phi_2\right) + m_{12}^2 \left(\Phi_1^\dagger\cdot \Phi_2+\Phi_2^\dagger\cdot \Phi_1\right)\nn\\
         &+ \lambda_1 \left(\Phi_1^\dagger\cdot \Phi_1\right)^2 + \lambda_2 \left(\Phi_2^\dagger\cdot \Phi_2\right)^2+\lambda_3 \left(\Phi_1^\dagger\cdot \Phi_1\right)\left(\Phi_2^\dagger\cdot \Phi_2\right)\nn\\
	 &+\lambda_4 \left(\Phi_1^\dagger\cdot \Phi_2\right)\left(\Phi_2^\dagger\cdot \Phi_1\right)+\frac{1}{2}\lambda_5\left[\left(\Phi_1^\dagger\cdot \Phi_2\right)^2+\left(\Phi_2^\dagger\cdot \Phi_1\right)^2\right]
\end{align}
One or both doublet(s) $\Phi_1 $ and $\Phi_2$ may acquire VEVs if $m_{ij}^2$ has one or two negative eigenvalues, and we write the doublets and their VEVs as
\begin{equation}
 \Phi_i=\begin{pmatrix}
               \Phi_i^+\\
               \Phi_i^0
              \end{pmatrix}
              \text{ and }
  \vev{\Phi_i}=\frac{1}{\sqrt{2}}\begin{pmatrix}
               0\\
               v_i
              \end{pmatrix},\,\text{ for }i=1,2.
\end{equation}
We then define the angle $\beta$ through the usual relation 
\begin{equation}
 \tan\beta=\frac{v_2}{v_1}\Leftrightarrow \left\{\begin{matrix} v_1=v\cos\beta\\v_2=v\sin\beta \end{matrix}\right.\quad
\end{equation}
where $v$ is defined by $v^2=v_1^2+v_2^2$. The 2HDM hence has seven free parameters which are
\begin{equation}
 \lambda_i\text{ (for }i\in\{1,2,3,4,5\}\text{)};\quad m_{12}^2;\quad\tan\beta.
\end{equation}

It is often more convenient to work in another basis -- the so-called Higgs basis $\{H_1,H_2\}$ -- where the neutral component of the doublet $H_1$ is aligned in field space with the 
total VEV $v$, with a rotation of angle $\beta$
\begin{equation}
 \left\{\begin{matrix}
         \Phi_1&=&H_1c_\beta-H_2s_\beta\\\Phi_2&=&H_1s_\beta+H_2c_\beta
        \end{matrix}
\right.\quad\Leftrightarrow\quad\left\{\begin{matrix}
                              H_1&=&\Phi_1c_\beta+H_2s_\beta\\H_2&=&-\Phi_1s_\beta+\Phi_2c_\beta
                             \end{matrix}
\right.
\end{equation}
We choose to write these two new doublets as
\begin{equation}
  H_1=\begin{pmatrix}
               H_1^+\\
               \frac{1}{\sqrt{2}}\big(v+H_1^0\big)
              \end{pmatrix},\quad\quad
 H_2=\begin{pmatrix}
               H_2^+\\
               \frac{1}{\sqrt{2}}H_2^0
              \end{pmatrix}.
\end{equation}

In this new basis, following the notation of \cite{Bernon:2015qea}, the potential can be written as
\begin{align}
 \vtree=&\,\,Y_1 \left(H_1^\dagger\cdot H_1\right) + Y_2 \left(H_2^\dagger\cdot H_2\right) + Y_3 \left(H_1^\dagger\cdot H_2+H_2^\dagger\cdot H_1\right)\nn\\
         &+ \frac{Z_1}{2} \left(H_1^\dagger\cdot H_1\right)^2 + \frac{Z_2}{2} \left(H_2^\dagger\cdot H_2\right)^2+Z_3 \left(H_1^\dagger\cdot H_1\right)\left(H_2^\dagger\cdot H_2\right)\nn\\
	 &+Z_4 \left(H_1^\dagger\cdot H_2\right)\left(H_2^\dagger\cdot H_1\right)+\frac{1}{2}Z_5\left[\left(H_1^\dagger\cdot H_2\right)^2+\left(H_2^\dagger\cdot H_1\right)^2\right]\nn\\
	 &+\left[Z_6\left(H_1^\dagger\cdot H_1\right)+Z_7\left(H_2^\dagger\cdot H_2\right)\right]\left[H_1^\dagger\cdot H_2+H_2^\dagger\cdot H_1\right].
\end{align}
The CP-even physical states are eigenstates of the mass matrix
\begin{align}
&\mathcal{M}_H^2
=\begin{pmatrix}
 Z_1v^2 & Z_6v^2 \\
 Z_6v^2 & m_A^2+Z_5v^2
 \end{pmatrix},\\
\text{where }&m_A^2=-\frac{2m_{12}^2}{s_{2\beta}}-\lambda_5v^2
\label{def_mA}
\end{align}
which is diagonalised with an angle $\alpha$, and are given by
\begin{equation}
\label{mass_eigenstates}
 \left\{\begin{matrix}
         h\,=(\sqrt{2}\real(\Phi_1^0)-v_1)s_\alpha+(\sqrt{2}\real(\Phi_2^0)-v_2)c_\alpha=\real(H_1^0)s_{\beta-\alpha}+\real(H_2^0)c_{\beta-\alpha}\\
         H=(\sqrt{2}\real(\Phi_1^0)-v_1)c_\alpha-(\sqrt{2}\real(\Phi_2^0)-v_2)s_\alpha=\real(H_1^0)c_{\beta-\alpha}-\real(H_2^0)s_{\beta-\alpha}
        \end{matrix}\right. .
\end{equation}
 
The \textit{alignment limit} is defined as the limit in which the neutral components of the Higgs-basis doublets are also mass eigenstates, or in other words, the limit in which one of the CP-even neutral scalar mass eigenstates is aligned with the VEV $v$. From eq. (\ref{mass_eigenstates}) we see that this can be realised in 
two ways: 
\begin{itemize}
 \item[(i)] $s_{\beta-\alpha}=0$ in which $H$ carries the VEV and is identified with the SM-like Higgs.
 \item[(ii)] $c_{\beta-\alpha}=0$ which means that $h$ is the SM-like Higgs.
\end{itemize}
We do not make any assumption on the size of the masses of the different scalars $i.e.$ we do not suppose that we are in the decoupling limit as well. Consequently, at tree-level we only require $Z_6v^2\rightarrow 0$, and hence with the expression of $Z_6$ derived in \cite{Bernon:2015qea}, we have
\begin{equation}
 Z_6\equiv-s_{2\beta}\left[\lambda_1c_\beta^2-\lambda_2s_\beta^2-\frac{1}{2}\lambda_{345}c_{2\beta}\right]=0
\end{equation}
where $\lambda_{345}\equiv\lambda_3+\lambda_4+\lambda_5$. The simplest, and $\tan\beta$-independent, way to fulfil this condition is to have
\begin{equation}
\label{tree_alignment}
 \lambda_1=\lambda_2=\frac{1}{2}\lambda_{345},
\end{equation}
which we will use in the following to constrain tree-level alignment. Also, we will require that the SM-like Higgs be the lightest mass eigenstate $h$ (case (ii) above), by ensuring that
\begin{equation}
 Z_1v^2<m_A^2+Z_5v^2
\end{equation}
This implies that $c_{\beta-\alpha}=0$, and thus, with the conventional choice that $\beta\in [0,\frac{\pi}{2}]$ and $|\alpha|\leq\frac{\pi}2$, we have that 
\begin{equation}
 \beta-\alpha=\frac\pi 2 \Rightarrow \alpha\in[-\frac\pi2,0].
\end{equation}
The constrain for tree-level alignment given in eq.~(\ref{tree_alignment}) reduces the number of free parameters of the model from seven to five, as two of the quartic couplings (eg. $\lambda_2$ and $\lambda_3$) can be found as a function of the three other ones. 

For most scans and figures presented below, we worked in the \textit{type-I} 2HDM if not indicated otherwise. However as the difference with \textit{type-II} comes from the couplings of the scalars to the down-type quarks and to the leptons which are light and give much smaller contributions to the lightest Higgs mass than the top quark, we do not expect large effects on our results (even for large $\tan \beta$, since the contributions typically involve the quark masses rather than just the couplings).  

\subsection{Renormalisation scale dependence of the Higgs mass computed with \SPheno}
The masses computed by \SPheno are pole masses, which should in principle not depend on the renormalisation scale at which they are computed. Evaluating the variation of the masses with the scale $Q$ hence provides a consistency check of our results and an estimate of the theoretical uncertainty as the variation of the two-loop masses with $Q$ corresponds roughly to the three-loop corrections. 
For this purpose, we have tuned the $\lambda_i$ couplings to ensure a two-loop Higgs mass of 125.09 GeV, at scale $Q=160$ GeV, together with tree-level alignment and find the following values (using \texttt{HiggsBounds} we have verified that this point in parameter space is not excluded by the current experimental constraints)
\begin{gather}
 \lambda_1=\lambda_2=0.0911, \qquad  \lambda_3=0.3322, \qquad \lambda_4=0.8000, \qquad \lambda_5=-0.9500\nn\\
 m_{12}^2=-50\,000 \text{ GeV}^2, \qquad  \tan\beta=50.
\end{gather}

\begin{figure}
\begin{center}
 \includegraphics[width=.85\textwidth]{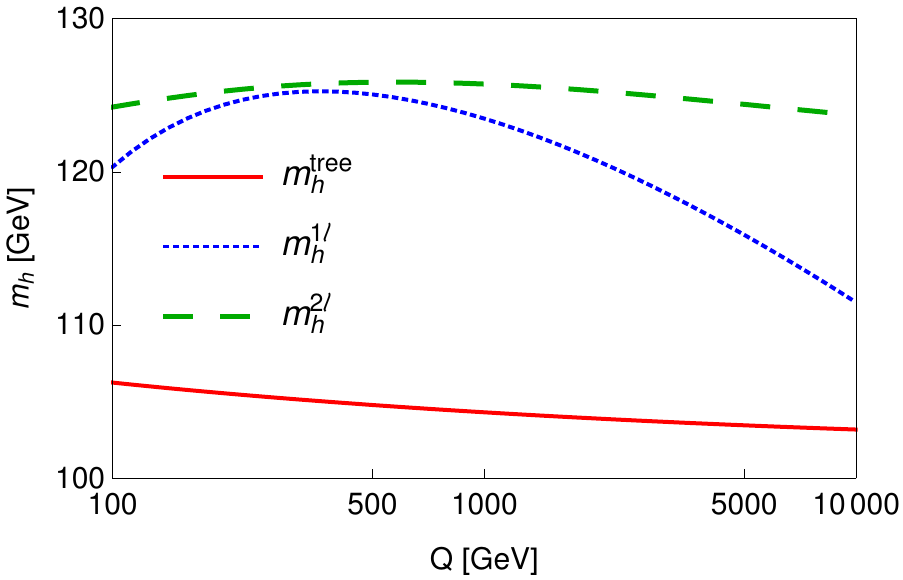}
 \caption{Lightest Higgs mass $m_h^2$ as a function of the renormalisation scale $Q$, considering only the running of SM parameters. \textit{Red curve: }tree-level; \textit{Blue dot-dashed curve: }one-loop order; \textit{Green dashed curve: }two-loop order.}
 \label{fig:Qdep}
\end{center}
\end{figure}

At first we consider that these inputs are then given to \SPheno as the value of the couplings at a scale $Q$ that we vary in the range $[100 \text{ GeV},10\,000 \text{ GeV}]$, and we only consider the running of SM parameters; we find the results shown in figure \ref{fig:Qdep} for the tree-level, one-loop and two-loop Higgs mass $m_h$. Since phenomenological analyses typically supply the quartic couplings without reference to a higher-energy theory or the scale where they are determined, this plot shows the importance of the choice of that scale. 
We have verified that the renormalisation scale dependence of $m_h|^\text{tree}$ is entirely due to the scale dependence of the Higgs VEV $v$, as the running of the quartic couplings is for the moment not applied. The renormalisation scale $Q$ is seen to have only a limited effect on the two-loop value of $m_h$ which varies of about 2 GeV on the range of scales considered here, while the one-loop result varies by about 15 GeV. Since the two-loop curve is so flat, this shows that most of the variation in the calculation of Higgs mass for the chosen quartics must come from variation of the Standard Model parameters, and that a two-loop calculation (rather than one-loop) is necessary not just for precision but also to ensure scale stability. 

\begin{figure}
\begin{center}
 \includegraphics[width=.85\textwidth]{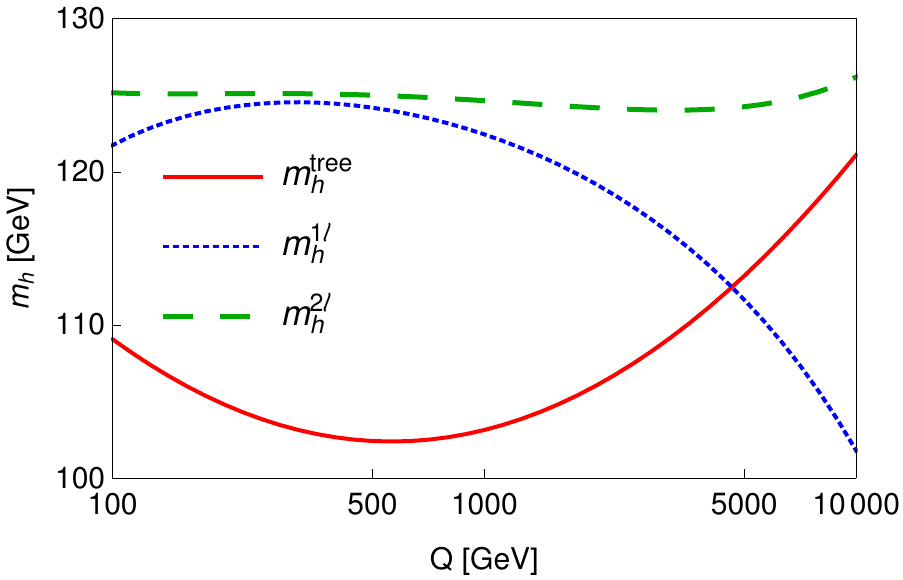}
 \caption{Lightest Higgs mass $m_h^2$ as a function of the renormalisation scale $Q$, taking into account the running of all parameters with the RGEs included in \SPheno. \textit{Red curve: }tree-level; \textit{Blue dot-dashed curve: }one-loop order; \textit{Green dashed curve: }two-loop order.}
 \label{fig:QdepallRGE}
\end{center}
\end{figure}

Using two-loop RGEs implemented in \SARAH/\SPheno, we can also include the evolution of the 2HDM parameters to obtain a more complete scale dependence of the masses, as shown in figure \ref{fig:QdepallRGE}. 
Once more, the two-loop value of Higgs mass depends less on the renormalisation scale than the tree-level or one-loop values. This smaller dependence of the two-loop Higgs mass on $Q$, compared with the one-loop mass, even for choices of parameters that give large loop corrections is a first verification of the validity of our new two-loop routines. In the following we will therefore work at a fixed scale $Q=m_t$, confident that the results will be for the most part independent of this choice.

\subsection{Quantum corrections to the alignment limit}
The relations defining the alignment limit in the beginning of this section are only valid at tree level 
and we expect them to receive corrections at one- and two-loop order, and in this section we will discuss the importance of these effects on the mixing angle of the neutral CP-even scalars $\alpha$.

\begin{figure}[tb]
 \begin{center}
  \includegraphics[width=.45\linewidth]{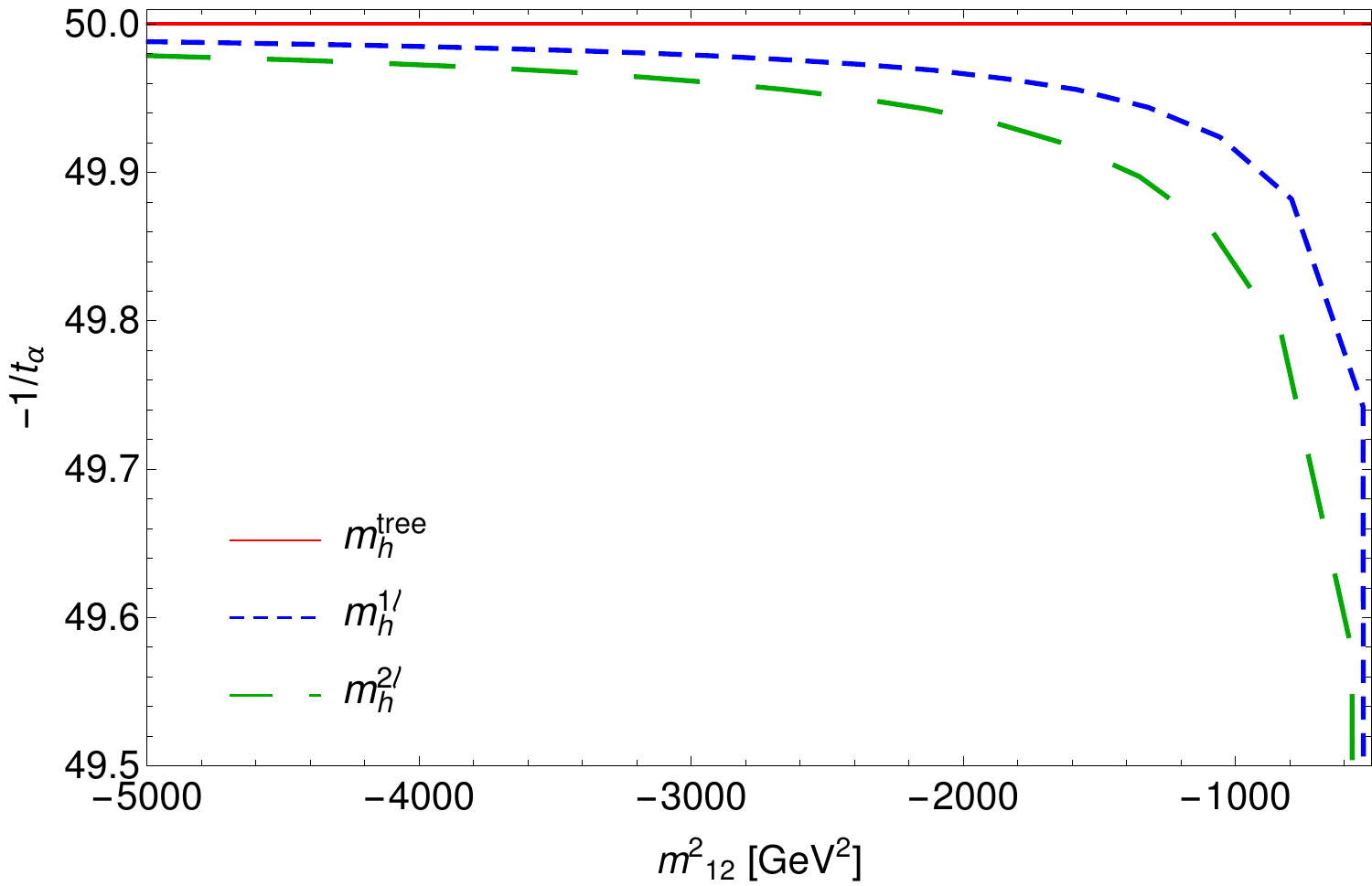}\quad
  \includegraphics[width=.45\linewidth]{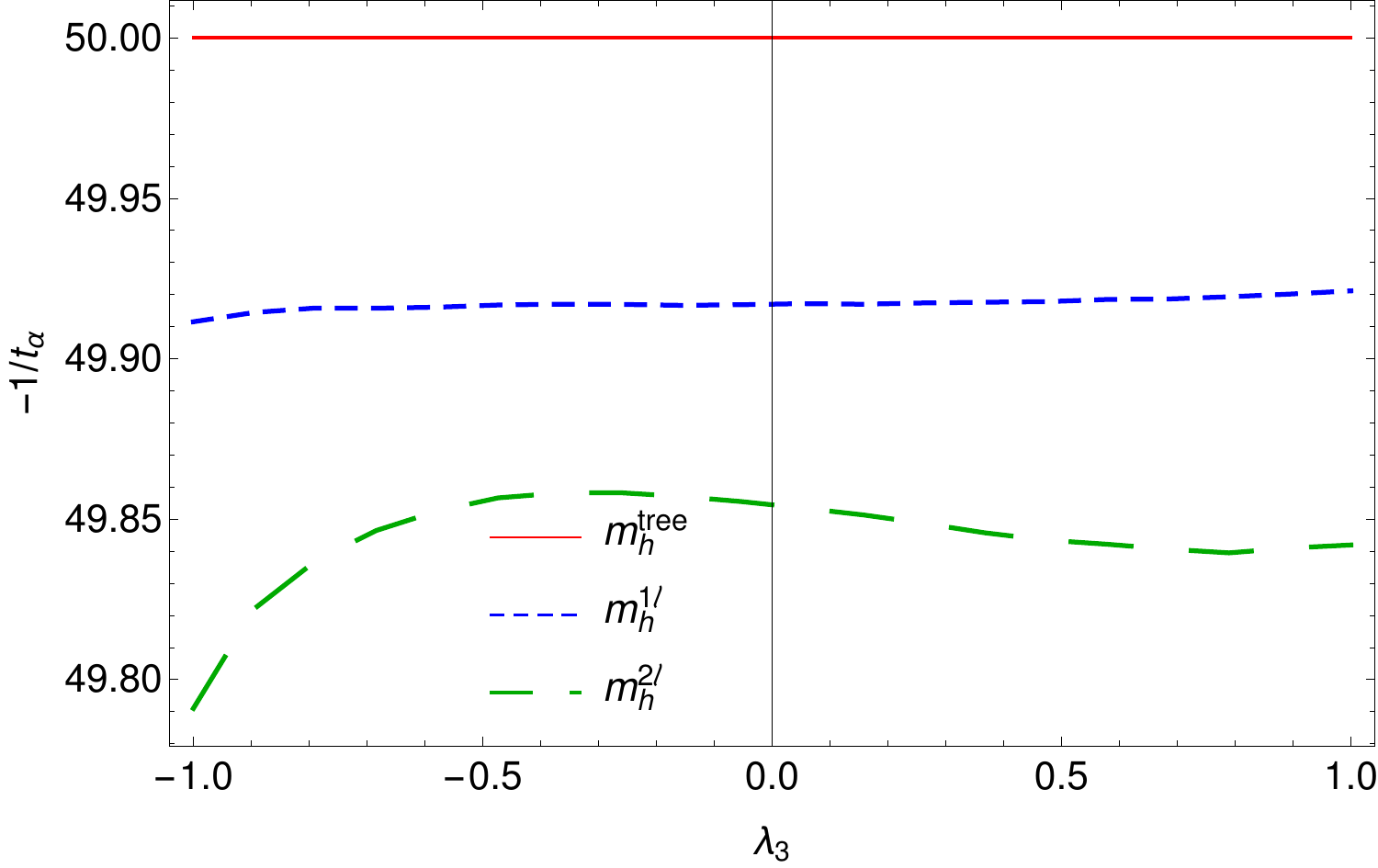}\\\vspace{.3cm}
  \includegraphics[width=.45\linewidth]{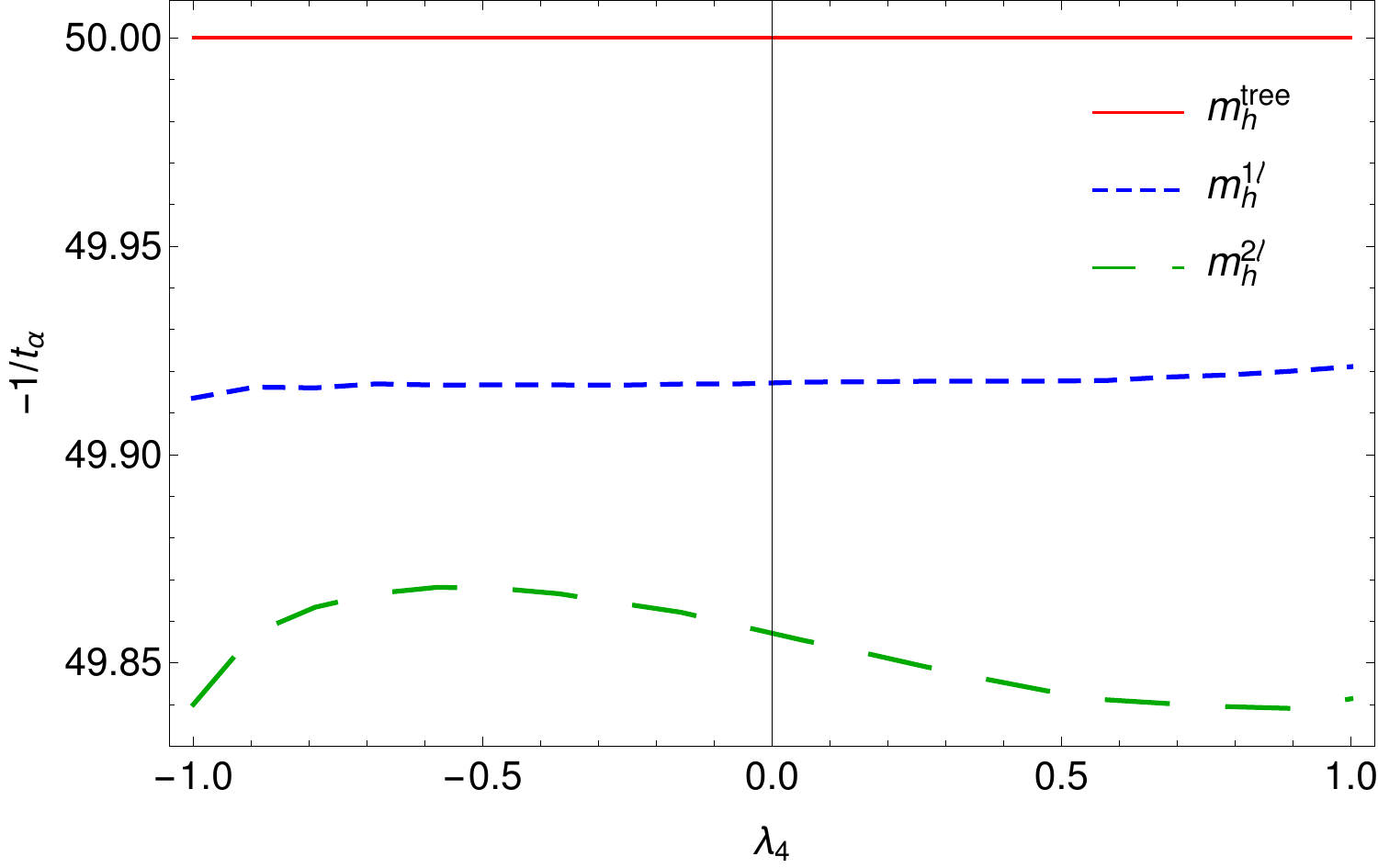}\quad
  \includegraphics[width=.45\linewidth]{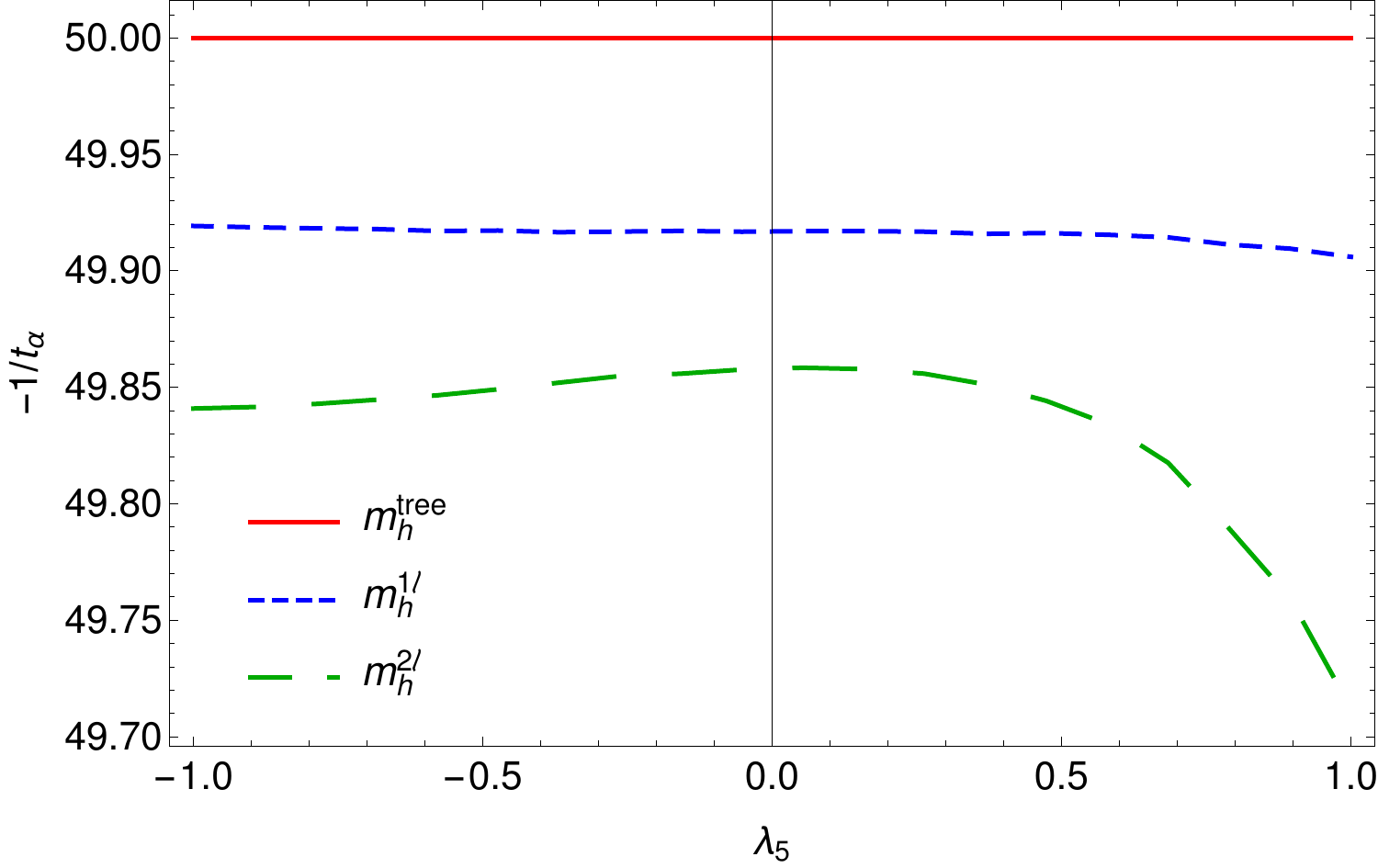}
  \end{center}\vspace{-15pt}
  \caption{$-1/t_\alpha$ as a function of the off-diagonal mass term $m_{12}^2$ \textit{(upper left)}, and of quartic couplings $\lambda_3$ \textit{(upper right)}, $\lambda_4$ \textit{(lower left)} and $\lambda_5$ \textit{(lower right)} at each order in perturbation theory. For each plot we vary the parameters as follows: we choose one parameter as the abscissa; the tree-level alignment condition $\lambda_1=\lambda_2=1/2\lambda_{345}$ plus the requirement that the Higgs mass is $125.09$ GeV fixes \emph{three} parameters, namely $\lambda_1, \lambda_2$ and either $\lambda_4$ for the bottom right plot or $\lambda_5$ for the other three; the remaining parameters are held fixed at values $\lambda_3=0.5,\,\lambda_4=0.5,\,m_{12}^2=-1000\text{ GeV}^2 $ (when they are not otherwise varying). All plots are for $\tan\beta=50$. \textit{Red curve: }tree-level; \textit{Blue dot-dashed curve: }one-loop order; \textit{Green dashed curve: }two-loop order.}
 \label{fig:corr_align}
 \end{figure}

Scanning over the different free couplings of the model -- $m_{12}^2$ and $\lambda_i$ ($i\in\{3,4,5\}$)  -- 
we compare the values of the CP-even Higgs mixing angle $\alpha$ at tree level, one-loop and two-loop order, as shown in figure  \ref{fig:corr_align}, and as expected, loop corrections cause deviations from the tree-level relation $t_\alpha=-1/t_\beta\Leftrightarrow c_{\beta-\alpha}=0$. The observations we can make from these plots are the following
\begin{itemize}
 \item[$(i)$] in the ranges of the parameters that we considered, the effect of loop corrections on the value of $\alpha$ is small, at most of the order of $1\%$;
 \item[$(ii)$] the one-loop corrections to $\alpha$ show very little dependence on the quartic couplings $\lambda_{i=3,4,5}$;
 \item[$(iii)$] it appears that for most parameter points, the two-loop corrections to $\alpha$ are of similar magnitude than the one-loop ones -- although somewhat smaller when $|\lambda_i|\lesssim 1$.
 \item[$(iv)$] for some parameter points however, the two-loop corrections to $\alpha$ become significantly larger than the one-loop corrections, see the lower right plot in figure \ref{fig:corr_align}. We have verified that this happens when one of the quartic couplings $\lambda_i$ becomes large (typically $|\lambda_i|\gtrsim 1$) -- in the plot mentioned above of $-1/t_\alpha$ as a function of $\lambda_5$ it is $\lambda_4$ that becomes smaller than $-1$ . We may suspect the large two-loop effects are due to a loss of perturbativity: this will be discussed in more detail in the next section.  
\end{itemize}

\subsection{Perturbativity constraints}
It is common in practice to use the physical scalar masses, the $\mathbb{Z}_2$ breaking parameter $m_{12}$  as well as the the angles $\alpha$,$\beta$ as input for the 2HDM in numerical studies. However, this input often hides that it corresponds to huge quartic couplings which spoil unitarity and the perturbative behaviour of the theory. Therefore, the constraints that all quartics must be smaller than $4\pi$ as well as the tree-level unitarity constraints \cite{Kanemura:1993hm,Akeroyd:2000wc,Horejsi:2005da} are applied to sort such points out. However, it was already shown in the SM that the limit of $\lambda<4\pi$ might be too weak \cite{Nierste:1995zx}. 

We now have all the machinery at hand to impose another constraint on the 2HDM model namely that the radiative corrections to the Higgs mass converge. We show here in one example that this can be a much stronger constraint than tree-level unitarity, while a more detailed analysis of this constraint on the parameter space of 2HDM models is left for future work. 

We consider here a point for type--II defined by \footnote{We used {\tt HiggsBounds} \cite{Bechtle:2008jh,Bechtle:2013wla} to check that this point passes all current collider limits.}
\begin{eqnarray}
& m_{H} = 593.6~\text{GeV},\quad m_A=535.2~\text{GeV},\quad m_{H^+} = 573.2~\text{GeV}& \nonumber \\
\label{eq:THDMp1}
&m_{12}^2 = - 165675~\text{GeV}^2,\quad \tan\alpha=-0.235,\quad \tan\beta=1.017 &
\end{eqnarray}
Since the masses are treated as pole-masses and only tree-level relations are used in the above work, no scale for the \MS parameters is given. On the other side, it is usually checked that the translation of these masses into quartic couplings results in parameters which are allowed by tree-level unitarity. However, this treatment implicitly assumes that one can define at each loop level suitable counter-terms to renormalise the Higgs sector in a way that the masses can be kept constant, and that this renormalisation converges. This is however not the case for the parameter point defined by eq.~(\ref{eq:THDMp1}) as one can see as follows. The given input translates into the following set of quartic interactions using the tree-level relations\footnote{Note, negative $\lambda_2$ is usually taken to be forbidden because the potential is unbounded from below. However, this only holds for the tree-level potential. If RGE effects are included, $\lambda_2$ becomes positive after a few hundred GeV of running \cite{Staub:2017ktc}.}:
\begin{equation}
\label{eq:THDMp1quartics}
 \lambda_1 = 2.831,\quad \lambda_2=-2.134,\quad \lambda_3=7.974,\quad \lambda_4=-0.660,\quad \lambda_5 = 0.753
\end{equation}
These fulfil  the tree-level unitarity constraints \cite{Kanemura:1993hm,Akeroyd:2000wc,Horejsi:2005da}. 

\begin{figure}[tb]
 \includegraphics[width=0.49\linewidth]{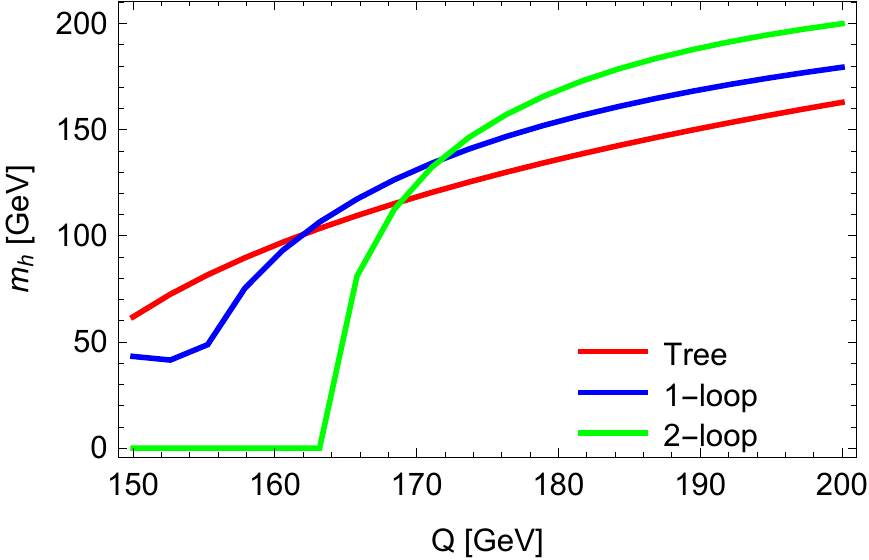} \hfill
 \includegraphics[width=0.49\linewidth]{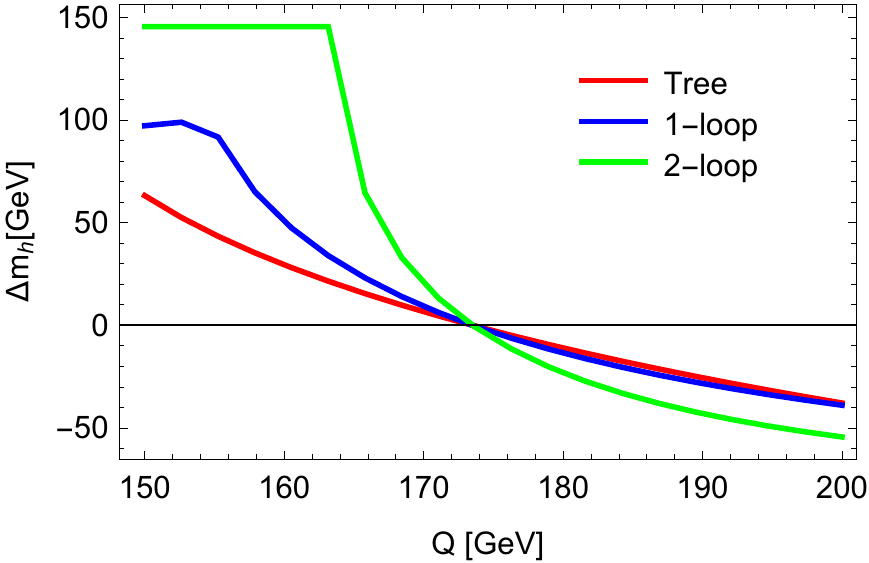}
 \caption{The dependence of the lightest scalar mass as function of the renormalisation scale $Q$ if the quartic couplings of eq.~(\ref{eq:THDMp1quartics}) are used as input 
 at the scale $Q=m_t$.}
 \label{fig:THDMqd}
\end{figure}

\begin{figure}[tb]
\centering
 \includegraphics[width=0.49\linewidth]{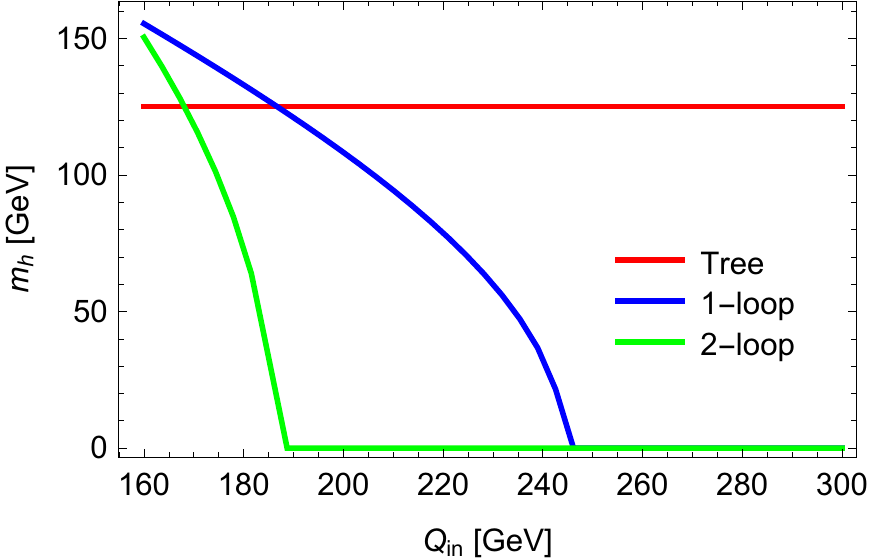} 
 \caption{The size of the one- and two-loop corrections of the lightest scalar mass as function of the scale $Q_{in}$ at which the input masses of eq.~(\ref{eq:THDMp1}) are translated 
 into quartic couplings. }
 \label{fig:THDMin}
\end{figure}

To check the perturbative behaviour, we show the scale dependence in figure~\ref{fig:THDMqd}. Here, we used the quartic couplings of eq.~(\ref{eq:THDMp1quartics}) as input and checked the scale dependence of the Higgs mass at different loop levels. 
For the evaluation of couplings to the considered scale, we used the two-loop RGEs calculated by \SARAH. One sees that the scale dependence increases with increasing loop-level. Of course, one might wonder if this is just an effect from our choice to define the quartic couplings at $Q=m_t$ as input. Therefore, we show in figure~\ref{fig:THDMin} the size of the loop corrections for different choices of our input scale $Q$. 
We see that the size of the loop corrections rapidly increases for $Q > m_t$ and the spread between one- and two-loop becomes even larger. Also choosing $Q\simeq~160\text{ GeV}$ where the mass at one and two-loop level seem to be roughly identical does not solve the problem: this is just a numerical coincidence and the scale dependence at two loops is even larger than at one loop.

\section{Georgi-Machacek Model}
\label{SEC:GM}

The Georgi-Machacek Model \cite{Georgi:1985nv}  extends the SM by one real scalar $SU(2)_L$-triplet $\eta$ with $Y=0$ and one complex scalar $SU(2)_L$-triplet $\chi$ with $Y=1$, which can be written as
\begin{equation} \label{eq:fieldsGM}
 \eta  = \frac{1}{\sqrt{2}} \, \left( \begin{array}{cc}
\eta^0 & - \sqrt{2} \left(\eta^-\right)^*\\
- \sqrt{2}  \eta^- & - \eta^0  \end{array} \right) \,, \quad 
 \chi  = \frac{1}{\sqrt{2}}  \, \left( \begin{array}{cc}
\chi^- & \sqrt{2} (\chi^0)^* \\
-\sqrt{2} \chi^{--} & -\chi^-  \end{array} \right) \,.
\end{equation}
A very compact form to write the Lagrangian in a $SU(2)_L \times SU(2)_R$ invariant form is to express the doublet and triplet scalars as
\begin{align}
\Phi =
\left(
\begin{array}{cc}
\phi^{0*}  & \phi^+     \\
 \phi^- & \phi^0      
\end{array}
\right)\,, \qquad
\quad \Delta =
\left(
\begin{array}{ccc}
 \chi^{0*} & \eta^+  & \chi^{++}   \\
 \chi^- & \eta^0  & \chi^+  \\
 \chi^{--} & \eta^{-}  & \chi^0   
\end{array}
\right)\,.
\end{align}
Here, $\phi$ are the components of the SM doublet. Using this notation, the scalar potential reads
\begin{align}
V(\Phi, \Delta) & = \mu_2^2 \mathrm{Tr} \Phi^\dagger \Phi + \frac{\mu_3^2}{2} \mathrm{Tr} \Delta^\dagger \Delta  + \lambda_1 \left[  \mathrm{Tr} \Phi^\dagger \Phi \right]^2 + \lambda_2  \mathrm{Tr} \Phi^\dagger \Phi \, \mathrm{Tr} \Delta^\dagger \Delta \nonumber \\
&  + \lambda_3 \mathrm{Tr} \Delta^\dagger \Delta \Delta^\dagger \Delta +   \lambda_4 \left[\mathrm{Tr} \Delta^\dagger \Delta \right]^2   - \lambda_5 \mathrm{Tr} \left( \Phi^\dagger \sigma^a \Phi \sigma^b  \right) \, \mathrm{Tr}  \left(\Delta^\dagger t^a \Delta t^b \right) \nonumber \\
&   - M_1  \mathrm{Tr} \left( \Phi^\dagger \tau^a \Phi \tau^b \right) (U \Delta U^\dagger)_{ab}\nonumber    - M_2 \mathrm{Tr} \left( \Delta^\dagger t^a \Delta t^b \right) (U \Delta U^\dagger)_{ab} \, ,
\end{align}
$\tau^a$ and $t^a$ are the $SU(2)$ generators for the doublet and triplet representations respectively, while $U$ is given for instance in \cite{Hartling:2014zca}. The triplets obtain VEVs as
\begin{equation} \label{eq:GM:VEVs}
\langle \eta \rangle = \frac{1}{\sqrt{2}} \, \left( \begin{array}{cc}
v_\eta & 0\\
0 & -v_\eta  \end{array} \right) \,, \quad 
\langle \chi \rangle =  \, \left( \begin{array}{cc}
0 & v_\chi \\
0 & 0  \end{array} \right) \, ,
\end{equation}
where the custodial symmetry enforces $v_\eta=v_\chi \equiv v_T $, and there are no tree-level contributions to the $\rho$ parameter. They further fulfil $v_\phi^2 + 8 v_T^2 = v^2,$ which allows us to define 
\begin{equation}
s_H = \sin \Theta_H = \frac{2\sqrt{2}v_T}{v}, \qquad c_H = \cos \Theta_H = \frac{v_\phi}{v} .
\end{equation}
The free parameters of the model are then 
\begin{equation}
\lambda_1 \dots \lambda_5\,,\quad M_1\,,\quad M_2\,,\quad s_H
\end{equation}
since $\mu_2^2, \mu_3^2$ can be eliminated by the tadpole equations. The physical eigenstates can be organised into representations of the custodial symmetry as a fiveplet (consisting of a doubly charged, singly charged a neutral CP-even scalar), a triplet (consisting of a singly charged and a CP-odd neutral scalar) and two CP-even singlets (where the Standard Model Higgs-like boson is the lighter of the two). Expressions for the triplet mass $m_3$, fiveplet mass $m_5$ and singlet masses are given in, for example, \cite{deFlorian:2016spz}. 

$m_h$, $s_H$, $m_5$ seem to be a suitable choice for the input parameters and can be traded for $\lambda_1$, $\lambda_5$ and $v_T$. In the following we shall do this using tree-level relations derived from those in \cite{deFlorian:2016spz}. However, the choice to use masses instead of couplings as input can have the danger that one enters a non-perturbative regime without recognising it as we already have pointed out for the 2HDM. We will discuss the importance of higher order corrections in general in this model in the following: in contrast for instance to the 2HDM, it is not possible to renormalise all mixing angles and masses on-shell in this model. One reason for this is that the masses of the five-plet are only exactly degenerate at tree-level but the custodial symmetry is not protected against loop effects \cite{Blasi:2017xmc}. Therefore, the number of mass parameters but also  of rotation angles is extended at the loop level: one needs three instead of two angles to diagonalise the loop corrected CP even mass matrix, and also the CP odd and charged Higgs mass matrix no longer share the same angle.
Therefore, an \MS renormalisation of the scalar sector is the natural option to check the impact of higher order corrections to the masses and angles. We give in Tab.~\ref{tab:GM} the loop corrected masses for all scalars for the parameter point $\lambda_2=\lambda_3=\lambda_4=0$, $m_5=1$~TeV and $s_H = 0.75$.\\
\begin{table}
\centering
\begin{tabular}{|c|c c c|}
\hline 
 & tree & one-loop & two-loop \\
\hline 
$m_{h_1}$~[GeV]     & 125.00  & 210.45  & $< 0$ \\
$m_{h_2}$~[GeV]     & 1000.00 & 950.56  & 916.96 \\
$m_{h_3}$~[GeV]     & 1054.67 & 975.20  & 954.03 \\
$m_{A_1}$~[GeV]     & 1049.31 & 998.41 & 896.13 \\
$m_{H^+_1}$~[GeV]   & 1000.00 & 950.80  & -\\
$m_{H^{+}_2}$~[GeV] & 1049.31 & 998.21 & -\\
$m_{H^{++}}$~[GeV]  & 1000.00 & 951.55  & -\\
\hline
\end{tabular} 
\caption{The scalar masses at tree- and loop-level for the parameter point $\lambda_2=\lambda_3=\lambda_4=0$, $m_5=1$~TeV and $s_H = 0.75$. The renormalisation scale was set to $m_5$.}
\label{tab:GM}
\end{table}
We see in these numbers that not only a mass splitting between the components of the 5 and 7-plets is induced at the one-loop level, but also that the loop corrections to the SM-like Higgs scalar can be huge. One can understand these large loop corrections for the chosen parameter point to some extent analytically: the one-loop corrections to the $(1,1)$-element of the CP even mass matrix are given in the effective potential in the limit $m_5 \gg v$ by
\begin{equation}
\Delta m_h \sim v^2 \frac{8 m_5^4 s_H^4}{9 \pi^2 v^4}.
\end{equation}
Thus, for large values of $m_5$ and/or $s_H$ one can expect huge corrections to the mass. Note, there are additional loop corrections to the off-diagonal elements of the scalar mass matrix which can have a significant impact on the masses. Therefore, one needs a full numerical calculation already at the one-loop level to obtain an accurate number for the SM-like Higgs mass. 

\begin{figure}[tb]
\begin{center}
\includegraphics[width=0.5\linewidth]{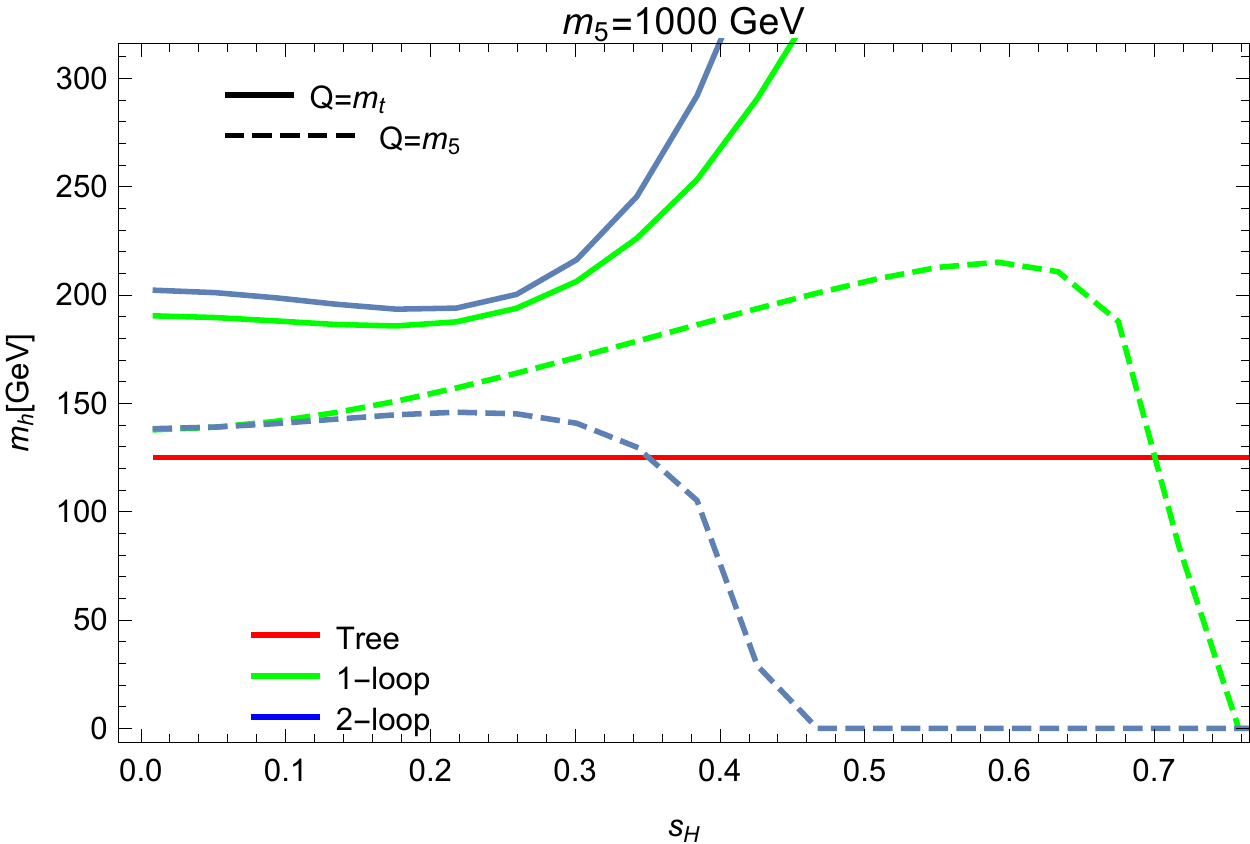}
\end{center}
\caption{The SM-like Higgs mass at tree-, one- and two-loop level for $m_5=1$~TeV and as function for $s_H$. The results for two different choices for the renormalisation $Q$ scale are shown.  }
\label{fig:GMq}
\end{figure}

Before we further investigate the loop corrections, we want to comment briefly on the choice for the renormalisation scale $Q$. In the SM, but also in other models like 2HDMs, it is suitable to set $Q=m_t$ to give a good convergence and ensure that there are no large logarithmic contributions from top loops. However, in the GM model the dominant loop corrections involve often scalar fields with masses $\propto m_5$. Therefore, the overall size of the loop corrections is usually smaller for $Q=m_5$ as one can see in figure~\ref{fig:GMq}.

\begin{figure}[tb]
\centering
\includegraphics[width=0.4\linewidth]{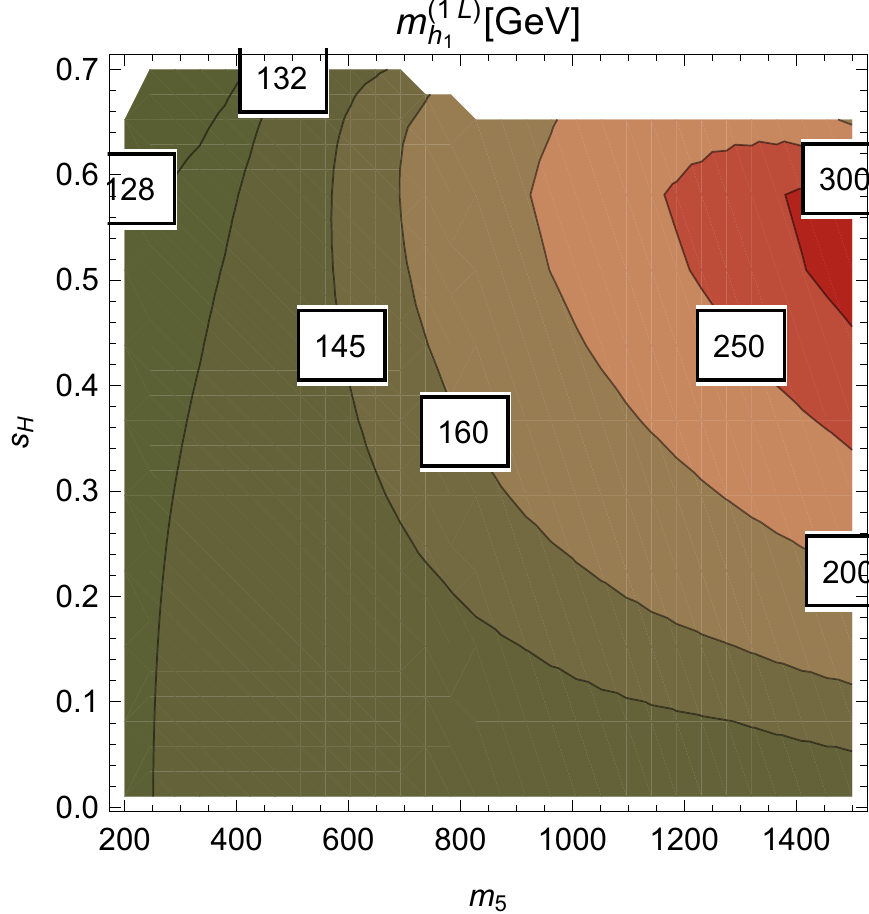}\quad
\includegraphics[width=0.4\linewidth]{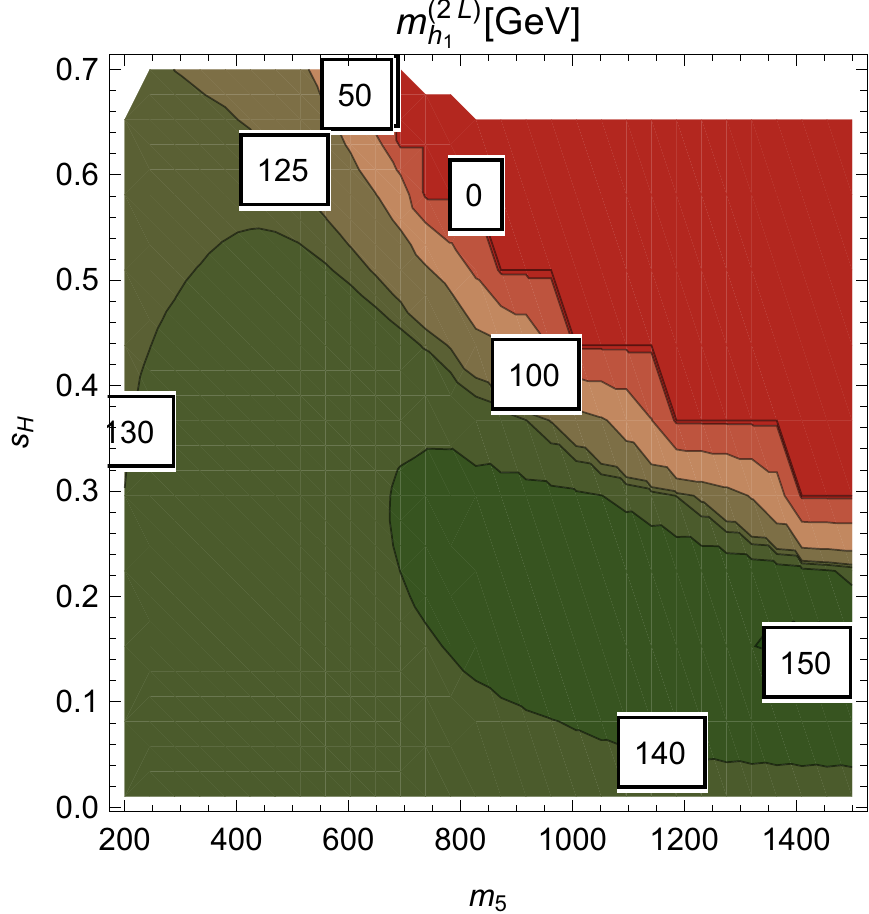} \\
\includegraphics[width=0.4\linewidth]{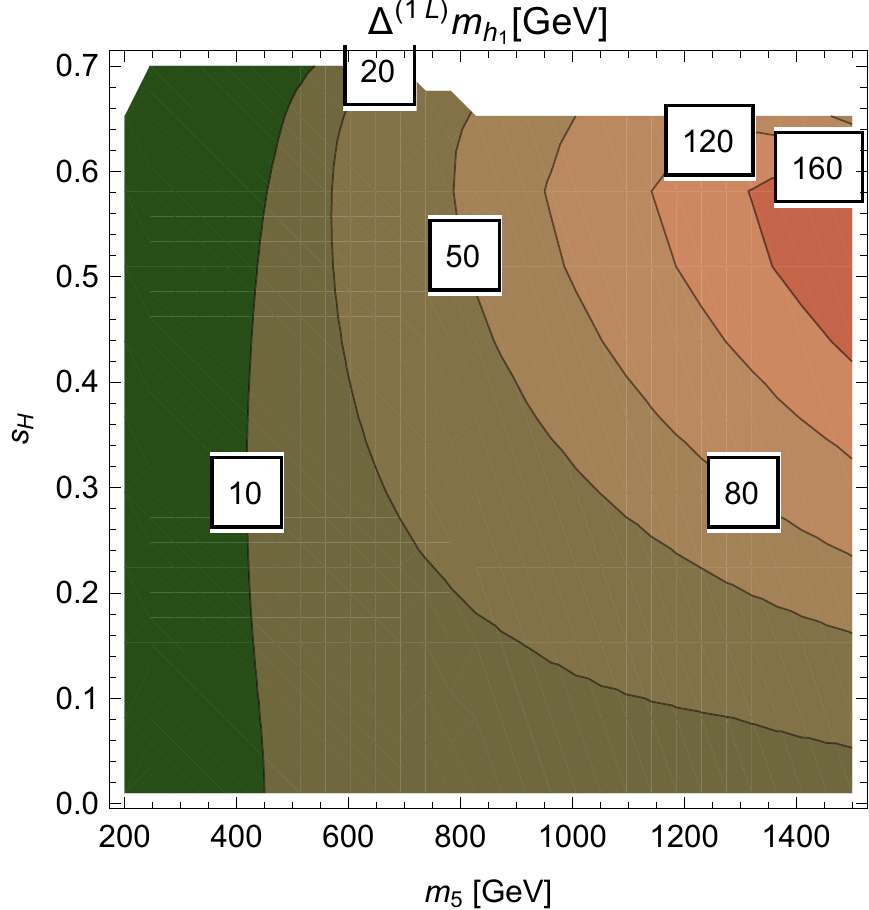}\quad
\includegraphics[width=0.4\linewidth]{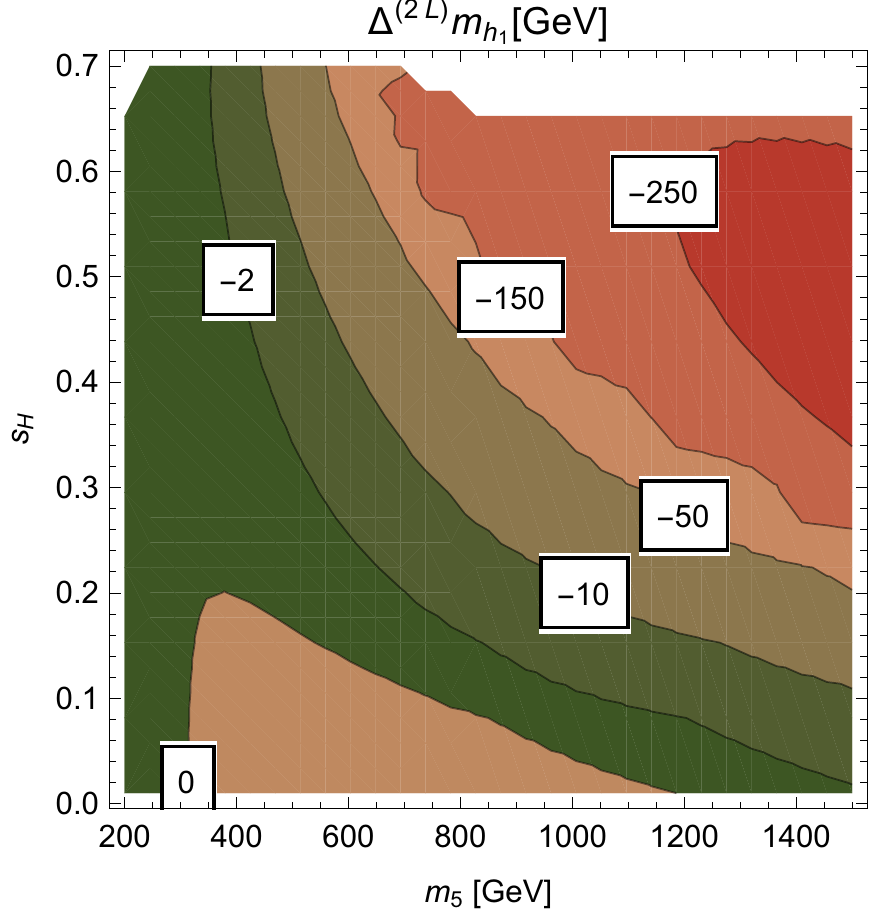} 
\caption{First row: absolute size of the  SM-like Higgs mass in the Georgi-Machacek model as function of $s_H$ and $m_5$ at including one- (left) and two-loop (right) corrections.
Second row: the size of the  one- (left) and two-loop (right) corrections. }
\label{fig:GM_MH1}
\end{figure}

We check now the Higgs mass in the $(m_5,s_H)$ plane proposed in \cite{deFlorian:2016spz} always  using $Q=m_5$. The other parameters are fixed in this plane to 
\begin{eqnarray*}
& m_h^{\rm tree} =  125~\text{GeV} \,,\quad M_1 =  \sqrt{2} \frac{s_H}{v} (m_5^2 + v^2)\,,\quad M_2 =  \frac16 M_1 & \\
& \lambda_3 = -0.1 \,,\quad \lambda_2 =  0.4 \frac{m_5}{1000~\text{GeV}} \,,\quad \lambda_4 =  0.2 &
\end{eqnarray*}
The light Higgs mass at the one- and two-loop level is shown in figure~\ref{fig:GM_MH1}. 
As expected, we see that the two-loop corrections are large for large $s_H$ and $m_5$. In order to further demonstrate this, we show in figure~\ref{fig:GM_MH1} also explicitly the size of the one- and two-loop corrections for all three CP-even scalars. 
\begin{figure}[tb]
\centering
\includegraphics[width=0.4\linewidth]{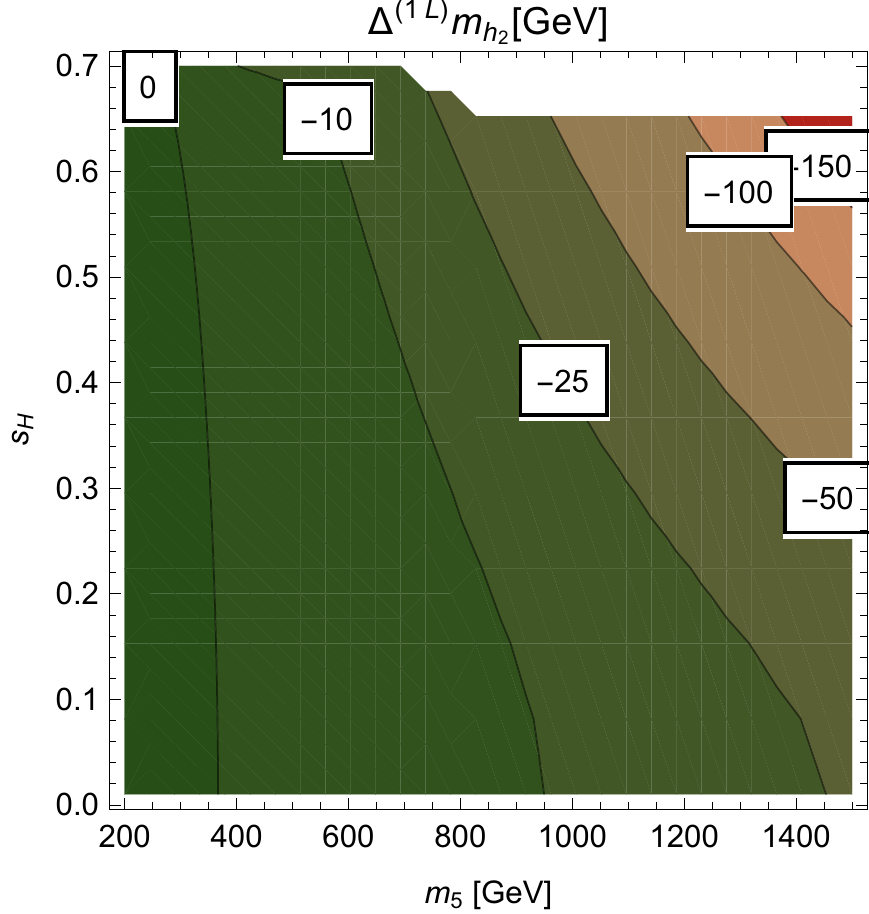}\quad
\includegraphics[width=0.4\linewidth]{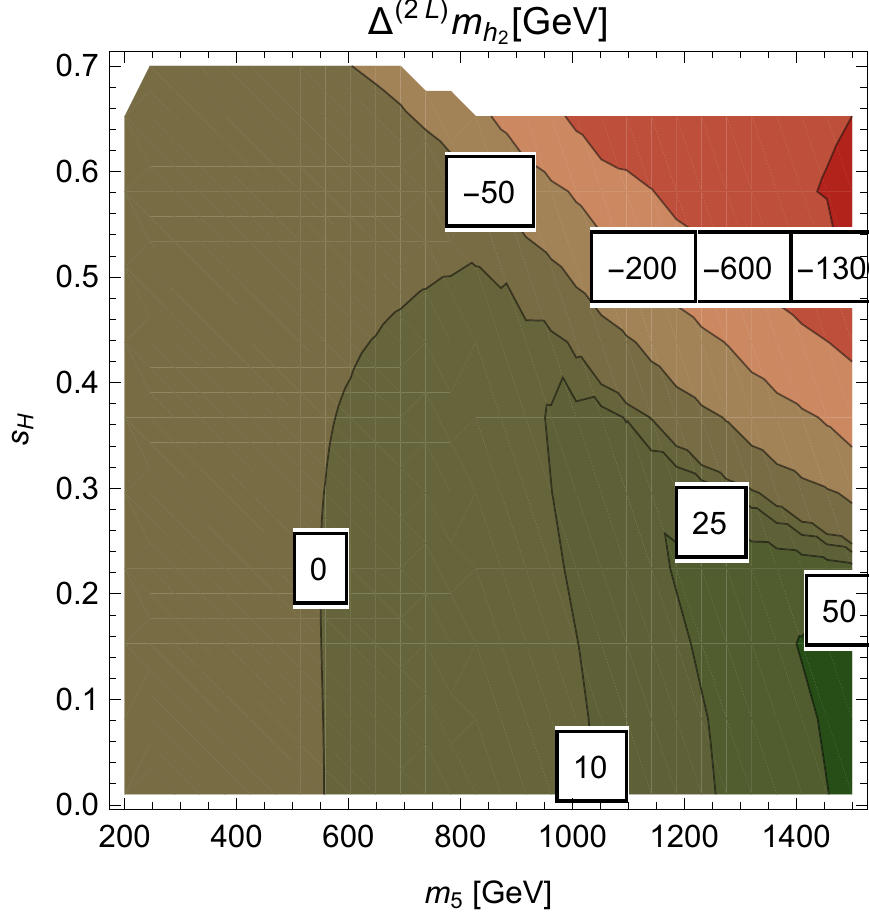}
\includegraphics[width=0.4\linewidth]{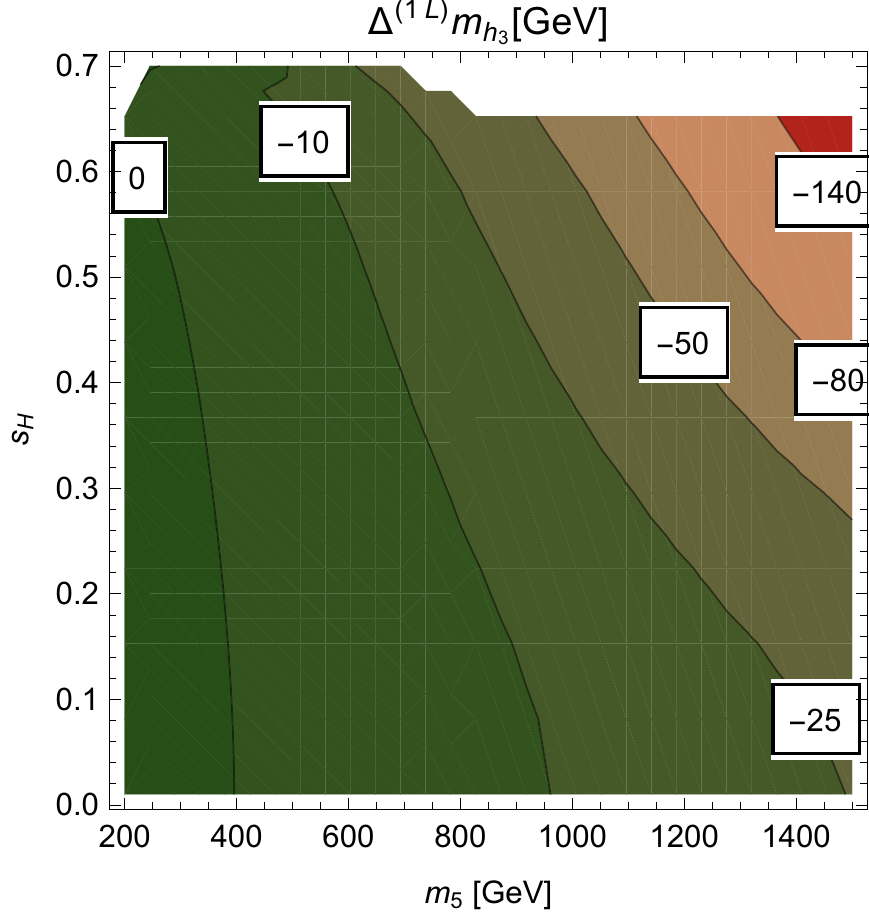}\quad
\includegraphics[width=0.4\linewidth]{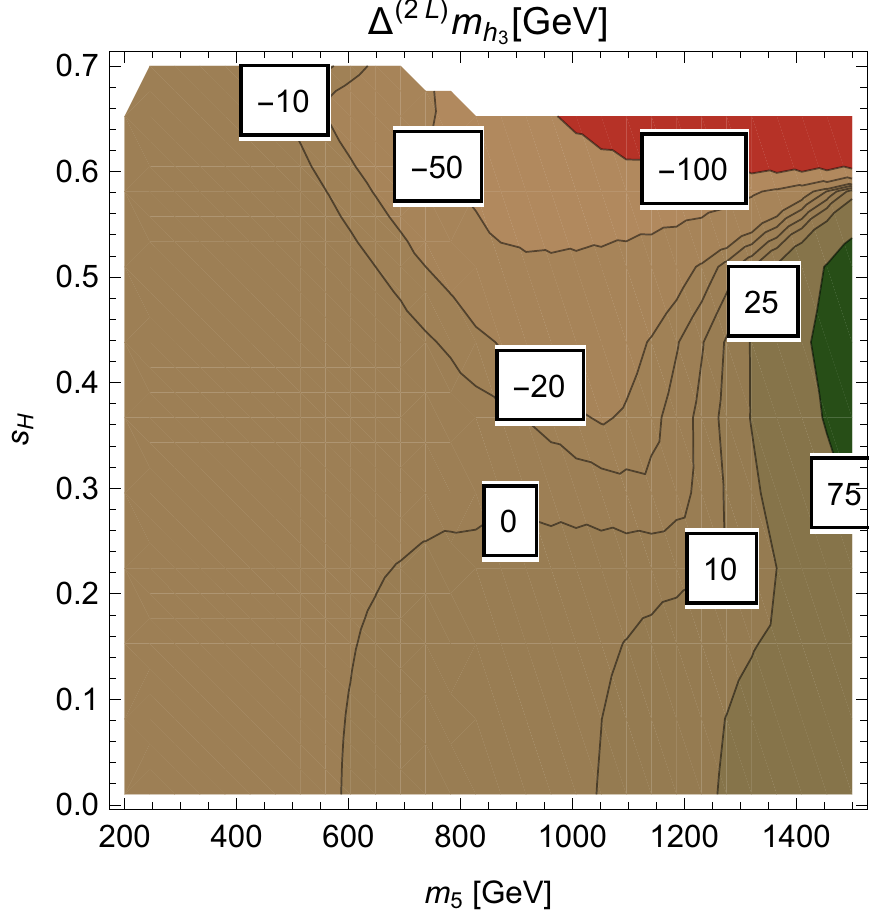}
\caption{The size of the  one- (left) and two-loop (right) corrections in the  $(s_H,m_5)$ plane for the second (first row) and third (second row) CP-even scalar.}
\label{fig:GM_DMH}
\end{figure}
We see that in the upper right corner in the $(s_H,m_5)$ plane the two-loop corrections are much larger than the one-loop ones and the Higgs can  even become tachyonic. For $m_5=1$~TeV, this already happens at $s_H > 0.5$, while for $m_5=1.5$~TeV the upper limit of $s_H$ is as low as $0.25$. For large $m_5$ this limit is much stronger than the one from perturbative unitarity of $VV\to VV$ scattering amplitudes which gives $s_H < \frac{667~\text{GeV}}{m_5}$ \cite{Logan:2015xpa}. Thus, even if it might still be possible to obtain the correct Higgs mass at two-loop level by adjusting the other input parameters or by absorbing finite corrections into counter-terms, the results in this parameter region should be taken with a lot of care. Most likely, they are meaningless. However, also for the other parameter regions with a reasonable hierarchy of the one- and two-loop corrections, one would need large adjustments in the input parameters to compensate for these loop corrections.   These changes would then reflect in the couplings and some decay widths of the 125-GeV scalar will deviate for large $s_H$ and/or $m_5$ clearly from the tree-level expectation.  Finally, one can also see in figure~\ref{fig:GM_DMH} that the loop corrections for the other scalars are sizeable and can shift the masses easily by tens to hundreds of GeV.

\section{Conclusions}
\label{SEC:Conclusions}

In this paper we have presented several varied results relating to the calculation of two-loop corrections to the Higgs mass in general models. Chief among these are:
\begin{enumerate}
\item We completed the basis of necessary loop functions for our on-shell solution, with a new expression for $\tilde{V}(0,x,y)$ given in appendix \ref{SEC:loopfn}. 
\item We extended the derivation of shifts to the tadpoles and Higgs mass from consistently solving the tadpole equations to include more general minimisation conditions, in particular allowing fermion masses to be directly dependent on the parameters (such as $\mu$ in the MSSM) with the expressions given in appendix \ref{APP:Consistent}.
\item We compared our results with those available for the Standard Model. In particular, this allowed a comparison within the same code of calculations in two different gauges, and we also found that the electroweak corrections are negligible, while those from momentum dependence are very small. 
\item We showed that our new computation does indeed remove the instabilities (sharp peaks in the Higgs mass for certain parameter choices) in the previous approach for supersymmetric models; however, the reader should  be aware that there are still some limitations when scalar masses in the loops become small compared to the renormalisation scale.
\item We explored the corrections to the mixing angle in the alignment limit in the Two Higgs Doublet Model using the \MS \emph{couplings} as inputs, and found that provided the quartic couplings are chosen to be small, the loop corrections are safely under control. 
\item We explored the 2HDM and Georgi-Machacek models with masses as physical inputs and using tree-level relations to obtain \MS couplings, as commonly done in the literature. We find that in most regions of the parameter space these lead to large quartic couplings, which rapidly lead to loss of control of the loop corrections. Perhaps surprisingly, this often occurs well before the couplings reach naive perturbativity bounds. 
\end{enumerate}
All of the shown results are available to the community with \SARAH version 4.12.0, and we hope that this contributes to an efficient and more precise study of many extensions of the SM; this should open the avenue to much future work. It would be particularly interesting to explore more carefully the relationship between on-shell and \MS calculations in non-supersymmetric models, to better understand how the divergent behaviour of the masses that we observe for the \MS scheme translates into differences in physical couplings -- or even possibly ruling out certain parameter regions of models as unphysical. 

For the technical program of generic Higgs mass computations, it would be very interesting to compute the corrections to the electroweak VEV and top Yukawa coupling to the same precision that we can achieve for the Higgs mass from \MS/\DR inputs. It would also be interesting to complete the set of contributions with those stemming from electroweak couplings, even if we showed that these must be very small in the case of the Standard Model.

\section*{Acknowledgements}

We thank Pietro Slavich for helpful discussions and many comments on the draft. 
JB and MDG acknowledge support from French state funds managed by the Agence Nationale de la Recherche (ANR), in the context of the LABEX
ILP (ANR-11-IDEX-0004-02, ANR-10-LABX-63), and MDG acknowledges
support from the ANR grant ``HiggsAutomator'' (ANR-15-CE31-0002). 
JB was supported by a scholarship from the Fondation CFM. 
FS is supported by ERC Recognition Award ERC-RA-0008 of the Helmholtz Association.

\appendix

\section{Loop functions}
\label{SEC:loopfn}

\subsection{$\tilde{V}(x,0,z,u)$}

One of the key functions of the basis set is $V(x,y,z,u)$. This is defined as
\begin{align}
V(x,y,z,u) \equiv& - \frac{\partial}{\partial y} U(x,y,z,u).
\end{align}
It is singular as $y\rightarrow 0$, so we define the regularised version:
\begin{align}
\ov{V} (x,y,z) &\equiv \lim_{u \rightarrow 0} \bigg[ V(x,u,y,z) - \frac{1}{s-x} \frac{\partial}{\partial u} I(u,y,z)\bigg].
\end{align}
On the other hand, we require a slightly \emph{different} regularised function:
\begin{align}
\tilde{V} (x,y,z) &\equiv \lim_{u \rightarrow 0} \bigg[ -V(x,u,y,z) + P_{SS}(y,z) B(u,x') \bigg].
\end{align}
For the case $x \ne 0$, we can simply extract the result at vanishing external momentum: 
\begin{align}
\lim_{s\rightarrow 0}\bar{V}(x,z,u) =&   \lim_{y\rightarrow 0} \bigg[ - U_0(x,y',z,u) -\frac{1}{x} P_{SS} (z,u) \blog y \bigg] \nn\\ 
=&  \frac{I(x,z,u) - I(0,z,u)}{x^2}= - \frac{1}{x} U_0 (x,0,z,u).
\end{align}
Then constructing $\tilde{V}$ gives
\begin{align}
\lim_{s\rightarrow 0}\tilde{V} (x,z,u) =& - \lim_{s\rightarrow 0} \bar{V}(x,z,u) - \frac{1}{x} \bigg[ R_{SS} (z,u) + P_{SS} (z,u) (\blog x - 1)\bigg] \nn\\
=& \frac{1}{x} \bigg[ U_0(x,0,z,u) + R_{SS} (z,u) + P_{SS} (z,u) (\blog x - 1)\bigg].
\end{align}

On the other hand, for $x\rightarrow 0$ -- or even for non-zero momentum -- a closed-form for either of these functions is not (until now) present in the literature, and its evaluation (using, for example {\tt TSIL}) is not straightforward. Indeed, in principle to evaluate the function $V(x,y,z,u)$ we should use the differential equations given in \cite{Martin:2003qz}, in this case
\begin{align}
\frac{\partial }{\partial y} U (x,y,z,u) =& k_{UU} U(x,y,z,u) + k_{UT1} T(x,z,u) + k_{UT2} T (u,x,z) + k_{UT2} T (z,x,u) \\
& + k_{US} \big[ S(x,z,u) - \frac{1}{2} (A(x) + A(z) + A(u) + I(y,z,u)) \big] + k_{UB} B(x,y) + k_U \nn\\
\equiv& k_{UU} U(x,y,z,u) + \Delta, \nn
\end{align}
where the coefficients of the loop functions are themselves functions of $s,x,y,z,u$. However, here we encounter the problem that several of these coefficients are actually singular as $y\rightarrow 0$ -- so we cannot simply substitute the right-hand side of the equation to determine $V(x,0,z,u)$! 

However, we can obtain such a closed-form expression by using the ansatz
\begin{align}
U(x,y,z,u) &= f_0(s;x,z,u) + f(s;x,z,u) A(y) + f_1(s;x,z,u) y + \mathcal{O}(y^2) \nn\\
f_0(s;x,z,u) &= U(x,0,z,u), \nn\\
k_U &= - \frac{1}{y} + k_{UU}^0 + \mathcal{O}(y) \nn\\
\Delta &= \frac{\Delta^{(-1)}}{y} + \Delta_l \blog y + \Delta^0 + ...
\end{align}
and substituting it into the above differential equation, to find $f$ and $f_1$:
\begin{align}
 f \blog y + f_1 + ... =& (- \frac{1}{y} + k_{UU}^0) \big(f_0(s;x,z,u) + f(s;x,z,u) A(y) + f_1(s;x,z,u) y \big) + \Delta \nn\\
=& - \frac{f_0}{y} - f \blog y + (f - f_1 + f_0 k_{UU}^0) + \frac{\Delta^{(-1)}}{y} + \Delta_l \blog y + \Delta^0 + ... \nn\\
\rightarrow \Delta^{(-1)} =& f_0 , \qquad f = \frac{1}{2} \Delta_l, \qquad f_1 =\frac{1}{2}  \bigg(f + \Delta^0  + \Delta^{(-1)} k_{UU}^0\bigg).
\end{align}
The form of $f$ must correspond to the singularity; indeed we have
\begin{align}
f(s;x,z,u) &= \frac{1}{s-x} P_{SS} (z,u). 
\end{align}
However, $f_1$ is more work; we eventually obtain in the limit $x\rightarrow0$ that we are interested in 
\begin{align}
\tilde{V} (0,z,u) =& \bigg( \frac{uz \log z/u}{(u-z)^3} + \frac{u+z}{2(u-z)^2}  \bigg) B(0,0) \nn\\
& +\frac{1}{s} \bigg[  \frac{2 A(u) A(z) + (u+z)^2 + 2 (u+z) I(z,u,0)}{2(u-z)^2} \bigg] \nn\\
& + \frac{1}{2} \bigg[ \mathcal{K}_{UT2} T(u,0,z) + \mathcal{K}_{UT3} T(z,0,u) + \mathcal{K}_{US} S(0,z,u) + \mathcal{K}_{U} \bigg]
\label{EQ:Vtilde}\end{align}
and
\begin{align}
f_1(s;0,z,u) =& \tilde{V} (0,z,u) - \frac{P_{SS}(z,u)}{s} \blog(-s), \nn\\
\bar{V}(0,z,u) =&   -f_1(s;0,z,u) + \frac{1}{s} R_{SS} (z,u) .
\end{align}
The coefficients defined in the above are
\begin{align}
\mathcal{K}_{UT2} =& - \frac{2uz (s+u-z)}{s(u-z)^3} \nn\\
\mathcal{K}_{UT3} =& \frac{2uz (s-u+z)}{s(u-z)^3} \nn\\
\mathcal{K}_{US} =& - \frac{2(u+z)}{s(u-z)^2} \nn\\
\mathcal{K}_{U} =& - \frac{(u+z)^2}{(u-z)^2 s} + \frac{5 (u+z)}{4(u-z)^2} .
\end{align}

If we then make our generalised effective potential expansion, we find 
\begin{align}
f_1(s;0,z,u) = & -\frac{P_{SS}(z,u) \blog (-s)}{s} \nn\\
&-\frac{\blog (-s)}{2(u-z)^3} \bigg[ u^2 - z^2 - 2 uz \log \frac{u}{z} \bigg] \nn\\
&+\frac{1}{4(u-z)^4} \bigg[ 5(u+z)^3 + 8uz I(u,z,0) \nn\\
& + 2z \blog z  (2u^2 - 11uz + z^2) + 2u \blog u( u^2 - 11uz + 2z^2) + 4uz (u+z) \blog u \blog z \bigg]  .
\end{align}
We do not need the limit when $u=z=0$ because in that case we have couplings $\lambda^{GGG}$. However, for $z=0$ or $u=0$ we do see that there is a smooth limit of the above.

Let us define
\begin{align}
f_1 = - \frac{P_{SS}(z,u)}{s} \blog(-s)+  f_1^\ell \blog (-s) + f_1^0.
\end{align}
We can then write
\begin{align}
\tilde{V} (0,z,u) =&  f_1^\ell \blog (-s) + f_1^0.
\end{align}
We have
\begin{align}
f_1^\ell (z,u) =& -\frac{1}{2(u-z)^3} \bigg[ u^2 - z^2 - 2 uz \log \frac{u}{z} \bigg] \nn\\
f_1^\ell (z,z) =& -\frac{1}{6z}, \qquad f_1^\ell (0,u) = - \frac{1}{2u}
\end{align}

If we now substitute in the standard expressions for $I(z,u,0)$ then we can simplify the above to
\begin{align}
f_1^0 (z,u)=  \frac{1}{4(u-z)^3} \bigg[& 5(u^2-z^2) + 2z \blog z  (2u-z +u \blog z) + 2u \blog u(u-2z -z \blog u) \nn\\
& -4uz \bigg(\mathrm{Li}_2 (1-z/u) -\mathrm{Li}_2 (1-u/z)\bigg) \bigg].
\end{align}
We can also it in a shorter but less symmetric form
\begin{align}
f_1^0 (z,u)= \frac{1}{4(u-z)^3} \bigg[& 5(u^2-z^2) + 2z \blog z  (2u-z +2u \blog u) + 2u \blog u(u-2z -2z \blog u) \nn\\
& - 8uz \mathrm{Li}_2 (1-z/u) \bigg]. 
\end{align}
We can also take the limits:
\begin{align}
f_1^0 (z,z) =& \frac{11 + 3 \blog z}{18 z} ,\qquad f_1^0(0,u) = \frac{5 + 2 \blog u}{4u}.
\end{align}

\subsubsection{Integral representation}

Our expression for $\tilde{V}$ actually lends itself to an interesting finite integral representation. We start with the definition
\begin{align}
\tilde{V}(0,z,u) \equiv& \lim_{y\rightarrow 0} \bigg[ -V(y,y,z,u) + B(s,y,y') P_{SS} (z,u) \bigg].
\end{align}
Then, using $C \equiv 16\pi^2 \frac{\mu^{2\epsilon}}{(2\pi)^{4-2\epsilon}}$ we have
\begin{align}
V(x,y,z,u) =& - \frac{\partial}{\partial y} U(x,y,z,u) \nn\\
=& - \frac{\partial}{\partial y} \lim_{\epsilon \rightarrow 0} \bigg[\mathbf{U} (x,y,z,u) + 1/2\epsilon^2 - 1/2\epsilon  - \mathbf{B} (x,y)/\epsilon \bigg] \\
=& \lim_{\epsilon \rightarrow 0} \bigg[ -\mathbf{U} (x,y',z,u) + \mathbf{B} (x,y')/\epsilon \bigg] \nn\\
P_{SS} (z,u) =& - B(0;z,u) = - \lim_{\epsilon \rightarrow 0} \bigg[ \mathbf{B} (0;z,u) - 1/\epsilon \bigg].
\end{align}
So then 
\begin{align}
V(x,y,z,u) =& \lim_{\epsilon \rightarrow 0} \bigg[ C^2 \int \int \frac{1}{k^2 + x} \frac{1}{((k-p)^2 + y)^2} \frac{1}{q^2+z} \frac{1}{(q+k-p)^2 +u}  + \mathbf{B}(x,y')/\epsilon \bigg] \nn\\
\tilde{V} (x,z,u) =& \lim_{y\rightarrow 0} \lim_{\epsilon \rightarrow 0} \bigg[  -\mathbf{V} (x,y,z,u) - \bB(x,y') ( \mathbf{P}(z,u) + \frac{1}{\epsilon}) + \bB(x,y') \mathbf{P}(z,u) + B(x,y') P_{SS} (z,u)\bigg] \nn\\ 
=& \lim_{y\rightarrow 0} \lim_{\epsilon \rightarrow 0} \bigg[  -\mathbf{V} (x,y,z,u)  + \bB(x,y') \mathbf{P}(z,u) \bigg] \nn\\
 =& \lim_{y\rightarrow 0} \lim_{\epsilon \rightarrow 0} \bigg[ -C^2 \int \int \frac{1}{k^2 + x} \frac{1}{((k-p)^2 + y)^2} \frac{1}{q^2+z} \frac{1}{(q+k-p)^2+ u} \nn\\
& +C^2 \int \int\frac{1}{k^2 + x}  \frac{1}{((k-p)^2 + y)^2}  \frac{1}{q^2 + z} \frac{1}{q^2 + u} \bigg] \nn\\
=& \lim_{\epsilon \rightarrow 0} \bigg[ C^2 \int \int \frac{1}{k^2 + x} \frac{1}{((k-p)^2 + y)^2} \frac{1}{q^2+z} \frac{2 q \cdot (k-p) + (k-p)^2}{(q^2 + u)( (q+k-p)^2+ u)} \bigg].
\end{align}
We can then integrate this expression. For the case $x\rightarrow 0$ we can simplify a little:
\begin{align}
\tilde{V} (0,z,u) =& \lim_{y\rightarrow 0}\lim_{\epsilon \rightarrow 0} \bigg[ C^2 \int \int \frac{1}{(k+p)^2} \frac{1}{(k^2 + y)^2} \frac{1}{q^2+z} \frac{2 q \cdot k + k^2}{(q^2 + u)( (q+k)^2+ u)} \bigg] \nn\\
\equiv& \lim_{\epsilon \rightarrow 0}\left(-\frac{1}{z-u} \mathbf{F}(z,u) \right) \nn\\
\mathbf{F} (z,u) \equiv& C^2 \int \int \frac{1}{(k+p)^2} \frac{1}{k^4} \frac{1}{q^2+z} \frac{2 q \cdot k + k^2}{ (q+k)^2+ u}.
\end{align}
This integral is finite; we have checked that explicitly performing the integral using {\tt TARCER} \cite{Mertig:1998vk} exactly  yields expression (\ref{EQ:Vtilde}).

\subsection{Limits of $M(0,y,0,u,v)$}

Here we shall give explicit limits of the M function:
\begin{equation}
\label{exprM0y0uv}
 M(0,y,0,u,v) = A_M(y,u,v) \llog(-s) + B_M(y,u,v)
\end{equation}
\begin{align}
 A_M(y,u,v)=&\frac{u\blog u}{(y-u)(u-v)}-\frac{y\blog y}{(y-u)(y-v)}-\frac{v\blog v}{(y-v)(u-v)},\\
 B_M(y,u,v) =& -(2 + \blog v) A_M(y,u,v) \nn\\
&  + \frac{u + v}{(y-u)(u-v)} \mathrm{Li}_2 (1 - u/v) - \frac{v + y}{(y-u)(y-v)} \mathrm{Li}_2( 1 - y/v) .
\end{align}
$A_M$ is symmetric on all three indices, and as we already have an expression for $M(0,0,0,0,v)$ \cite{Braathen:2016cqe}, and as $M(0,0,0,u,0)$ or $M(0,y,0,0,0)$ have prefactor $\lambda^{GGG}$, we only need to consider the following cases
\begin{align}
A_M(0,u,v) =& \frac{\blog(v/u)}{u-v} \nn\\
B_M(0,u,v) =& -(2 + \blog v)A_M(0,u,v) - \frac{\pi^2}{6u} - \frac{(u+v) \mathrm{Li}_2 (1-u/v)}{u(u-v)} \nn\\
B_M(y,u,0)=& \frac{\log u/y \big[ 4 + \blog u + \blog y \big]}{2(u-y)}\nn\\
B_M(y,y,0)=& \frac{2 + \blog y}{y} \nn\\
A_M(y,y,v) =& \frac{v \log y/v}{(y-v)^2} - \frac{1}{y-v} \nn\\
A_M(y,y,0) =& - \frac{1}{y} \nn\\
B_M(y,y,v) =& -(2 + \blog v)A_M(y,y,v) + \frac{1}{(y-v)^2} \bigg[ (v+y) \log y/v + 2 v \mathrm{Li}_2 (1-y/v) \bigg]\nn\\
A_M(y,y,y) =& - \frac{1}{2y} \nn\\
B_M(y,y,y) =& \frac{1}{2y}(3+ \blog (y)) = -(2 + \blog y)A_M (y,y,y) + \frac{1}{2y}\nn\\
B_M(y,u,u) =& -(2 + \blog u)A_M(y,u,u) + \frac{2}{u-y} - \frac{(u+y) \mathrm{Li}_2 (1-y/u)}{(u-y)^2} 
\end{align}

\section{Consistent solution of the tadpole equations with shifts to fermion masses}
\label{APP:Consistent}

Here we give the two-loop shifts to the tadpoles and self-energies due to shifts in fermion masses when we solve the tadpole equations consistently.  

We denote the undiagonalised fermion mass matrix as $m^{IJ}$. The mass-squared matrix is defined \cite{Martin:2001vx} as 
\begin{align}
 \du{(m^2)}{I}{J}&=m_{IK}^*m^{KJ},
\end{align}
and is diagonalised by a unitary matrix $N$ defined such that 
\begin{align}
 m_I^2\du{\delta}{I}{J}=&\nm{K}{I}\cnm{L}{J}\du{(m^2)}{K}{L}, \qquad M^{IJ}\equiv \cnm{K}{I}\cnm{L}{J}m^{KL} \nn\\
\rightarrow M^{IK} M_{JK} =& m_I^2\du{\delta}{I}{J}.
\end{align}
Then if the tree-level matrices depend on some parameters $\{x_i\}$ for which we solve the tadpole equations as in equation (\ref{EQ:tadpoleshift}) we have
\begin{align}
\delta M^{IJ} =& \cnm{K}{I} \frac{\partial m^{KL}}{\partial x_k} \cnm{L}{J} c_{kl} \frac{\partial \Delta V^{(1)}}{\partial \phi_l^0}.
\end{align}
Then the  shift to the fermion contribution to the tadpole is
\begin{align}
 \delta^{(2)}\bigg(\left.\frac{\partial\vone_F}{\partial\phi_r^0}\right|_{\varphi=v}\bigg)=&-R_{rp}\re[y^{KLp}\delta M^*_{KL}]\big(A( m_K^2)+A( m_L^2)\big)\nn\\
 &-2R_{rp}\re[Y^{IJp}  M^*_{JK}]\big(\delta M^{KL} M^*_{IL}+\delta M^*_{IL} M^{KL}\big)P_{SS}( m_I^2, m_K^2),
\end{align}
while the shift to the scalar self-energy is
\begin{align}
 \delta\Pi^{(2),F}_{ij}=&-2\re[y^{KLi}y_{K'Lj}]\big( M^*_{KJ}\delta M^{K'J}+\delta M^*_{KJ}\bar M^{K'J}\big)\nn\\
&\times\bigg[ P_{SS}(m_K^2,m_{K'}^2)-B(m_{K'}^2, m_L^2)-\left(m_K^2+ m_{K'}^2-s\right)\,C(s,s,0, m_K^2, m_L^2, m_{K'}^2) \bigg]\nn\\
                        &+4\re[y^{KLi}y^{K'L'j}\delta M^*_{KK'} M^*_{LL'}]B( m_K^2, m_L^2)\nn\\
                        &+4\re[y^{KLi}y^{K'L'j}M^*_{IK'} M^*_{LL'}]\big( M^*_{KJ}\delta M^{IJ}+\delta M^*_{KJ} M^{IJ}\big)C(s,s,0, m_K^2, m_L^2, m_I^2).
\end{align}

To illustrate this, consider the MSSM, where the tadpole equations read
\begin{align}
(|\mu|^2 + m_{H_u}^2) v_u - B_\mu v_d + \frac{1}{8} (g_Y^2 + g_2^2) (v_u^2 - v_d^2) v_u =& - \frac{\partial \Delta V}{\partial v_u} \nn\\
(|\mu|^2 + m_{H_d}^2) v_d - B_\mu v_u - \frac{1}{8} (g_Y^2 + g_2^2) (v_u^2 - v_d^2) v_d =& - \frac{\partial \Delta V}{\partial v_d}.
\end{align}
Solving for $|\mu|^2$ we have 
\begin{align}
|\mu|^2 =& -\frac{M_Z^2}{2} + \frac{1}{c_{2\beta}} \bigg[ m_{H_u}^2 s_\beta^2 - m_{H_d}^2 c_\beta^2 + \frac{1}{v} s_\beta \frac{\partial \Delta V}{\partial v_u} - \frac{1}{v} c_\beta \frac{\partial \Delta V}{\partial v_d} \bigg],
\end{align}
so we have 
\begin{align}
\delta \mu =& \frac{1}{2\mu^* v c_{2\beta}} \bigg[  s_\beta \frac{\partial \Delta V}{\partial v_u} - c_\beta \frac{\partial \Delta V}{\partial v_d} \bigg].
\end{align}
This in turn will lead to a shift in the neutralino and chargino masses, which lead to a shift to the two-loop tadpoles.

\bibliographystyle{utphys}
\bibliography{NonSusy}

\end{document}